\documentclass[12pt]{article}

\usepackage{scicite}
\usepackage{graphicx}
\usepackage{amsmath,amssymb,amsfonts}
\usepackage{xcolor}

\usepackage{times}

\def\bk{{\mathbf{k}}}
\def\bh{\mathbf{h}}

\newenvironment{sciabstract}{%
\begin{quote} \bf}
{\end{quote}}

\title{The quantum metric of electrons with \\ spin-momentum locking}

\author
{Giacomo Sala,$^{1,\ast}$ Maria Teresa Mercaldo,$^{2}$ Klevis Domi,$^{1}$ Stefano Gariglio,$^{1}$ \\
Mario Cuoco,$^{3}$ Carmine Ortix,$^{2,\ast}$ Andrea D. Caviglia$^{1,\ast}$
\\
\normalsize{$^{1}$Department of Quantum Matter Physics, University of Geneva, Geneva, Switzerland}\\
\normalsize{$^{2}$Dipartimento di Fisica ‘E. R. Caianiello', Università di Salerno, Fisciano, Italy}\\
\normalsize{$^{3}$CNR-SPIN c/o Università di Salerno, Fisciano, Italy}\\
\\
\normalsize{$^\ast$To whom correspondence should be addressed;} \\ 
\normalsize{E-mail:  giacomo.sala@unige.ch, cortix@unisa.it, andrea.caviglia@unige.ch}
}

\date{}

\begin{document} 


\baselineskip24pt


\maketitle


\begin{sciabstract}
Quantum materials are characterized by electromagnetic responses intrinsically linked to the geometry and topology of electronic wavefunctions, encoded in the quantum metric and Berry curvature. Whereas Berry curvature-mediated transport effects have been identified in several magnetic and nonmagnetic systems, quantum metric-induced transport phenomena remain limited to topological antiferromagnets. Here we show that spin-momentum locking -- a general characteristic of the electronic states at surfaces and interfaces of spin-orbit coupled materials --  leads to a finite quantum metric. This metric activates a nonlinear in-plane magnetoresistance that we measure and electrically control in 111-oriented LaAlO$_3$/SrTiO$_3$ interfaces. These findings demonstrate the existence of quantum metric effects in a vast class of materials and enable previously unexplored strategies to design functionalities based on quantum geometry.
\end{sciabstract}

\noindent Nonlinear electronic responses can reveal physical properties that are out of the reach of linear probes \cite{tokura2018}. For example, the anomalous Hall effect senses the Berry curvature of magnetic materials~\cite{Xiao2010,Nagaosa2010}. In comparison, its nonlinear, second-order counterpart can exist even in the presence of time-reversal symmetry in noncentrosymmetric and nonmagnetic materials and provides information on closely-related geometric quantities: the Berry curvature dipole~\cite{Sodemann2015,Ma2019a,Lesne2022a,Du2021,Ortix2021} and the Berry curvature triple~\cite{Makushko2023}. 

Nonlinear electronic responses can manifest themselves as a nonreciprocal magnetoresistance, which was conventionally associated with disorder \cite{tokura2018}. However, recent works have identified an intrinsic nonlinear magnetoresistance driven by the quantum metric of electronic wavefunctions~\cite{Kaplan2022,Lahiri2023}. The quantum metric $g = \mathfrak{R}(G)$ corresponds to the real part of the quantum geometric tensor $G$ and, together with the Berry curvature $\Omega = -2\mathfrak{I}(G)$, defines the  geometrical and topological characteristics of quantum systems \cite{Provost1980,Cheng2010,Torma2023}. 
Despite the profound implications of the quantum metric for several physical phenomena~\cite{Torma2023,Resta2011,Peotta2015,Rossi2021,Huhtinen2022}, 
its experimental observations remain scarce \cite{Tan2019a,Gianfrate2020,Yu2020,Ren2021,Liao2021,Tian2023,Yi2023} and, in metallic systems, limited to topological antiferromagnets~\cite{Gao2023,Wang2023c,Han2024a}.

Here, we theoretically predict and experimentally demonstrate that a condensed-matter phenomenon as fundamental \cite{Li2014,Bihlmayer2022} and technologically relevant \cite{Manipatruni2019,Noel2020} as the spin-momentum locking of electronic states at surfaces and interfaces of spin-orbit coupled materials 
is endowed with a finite quantum metric. This intrinsic geometrical feature 
activates a nonlinear in-plane magnetoresistance that we use to provide evidence of the quantum metric at 111-oriented LaAlO$_3$/SrTiO$_3$ interfaces. 
Our findings and the abundance of materials with spin-momentum locked electronic bands broaden the significance of the quantum metric in condensed matter and opens up a whole line of possibilities to explore quantum geometry in a variety of systems \cite{Torma2023}.

\paragraph*{Quantum metric of spin-momentum locked electronic bands}
The direct relation between the spin-momentum locking and the quantum metric can be found by considering the general case of a single pair of Kramers related bands with parabolic dispersion that are shifted in momentum space by a Rashba-like linear term $\alpha_{\textrm{R}}(\mathbf{z}\times\mathbf{k})\cdot\pmb{\sigma}$~\cite{Bihlmayer2022}. Here, $\mathbf{z}$ is the direction normal to the surface or interface of interest and $\alpha_{\textrm{R}}$ is the Rashba parameter. The linear coupling between the crystalline momentum $\mathbf{k}$ and the electron spins $\pmb{\sigma}$ leads to helical spin textures (Fig. 1A) that are characterized by the complete absence of the Berry curvature~\cite{Lesne2022a}. Nevertheless, we find here that the quantum metric of spin-locked electronic bands is generally finite. As shown in Fig. 1B, the metric diagonal components $g_{\textrm{xx}}$ and $g_{\textrm{yy}}$ vanish only along the $k_{\textrm{y}} = 0$ and $k_{\textrm{x}} = 0$ lines, respectively, and possess a singular point at the time-reversal invariant momentum ${\bf k} = 0$. The band-energy normalized quantum metrics (BNQM, also known as Berry connection polarizabilities) of the spin-split bands cancel each other at the same crystal momentum. However, in the region of momenta populated by a single spin band the BNQM is finite. This implies that the dipole density components $\Lambda_{\textrm{xxx}}$ and $\Lambda_{\textrm{yyy}}$ of the BNQM governing the longitudinal nonlinear conductivities feature alternating positive and negative hotspots with a characteristic quadrupolar profile in momentum space (Fig. 1C and \cite{suppScience}). In the absence of magnetic fields, i.e., in time-reversal invariant conditions, the integrated dipole components vanish and so does the nonlinear response associated with the quantum metric \cite{Kaplan2022}.

\begin{figure*}[!tb]
\centering
\includegraphics{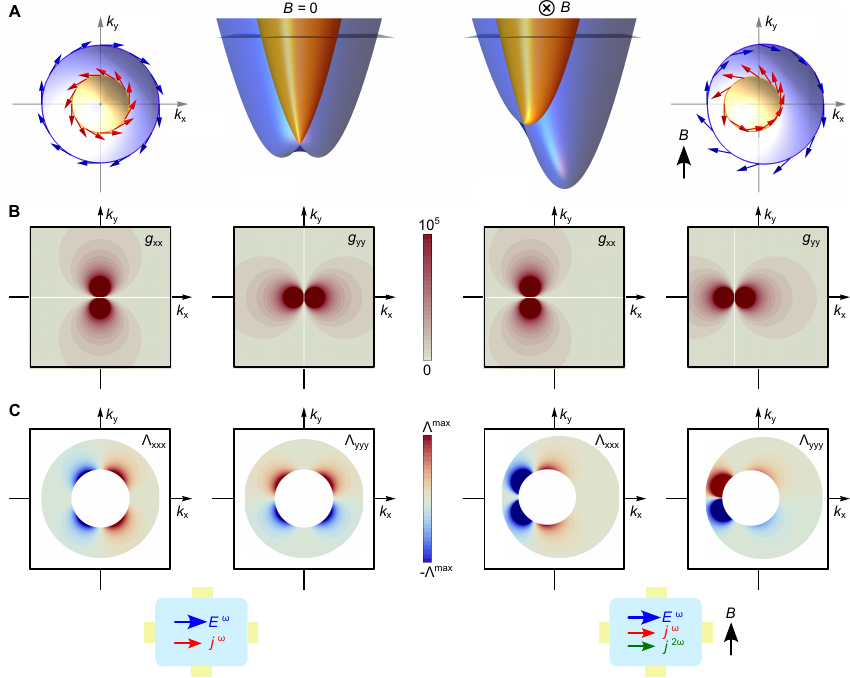}
\caption{\textbf{Quantum geometry of 2D Rashba bands.} (\textbf{A}) Spin-locked electronic bands induced by the 2D Rashba effect in the absence (left) and presence (right) of a planar magnetic field $B=0.1 E_0$, with $E_0$ a reference energy \cite{suppScience}. (\textbf{B}) Reciprocal space maps of the diagonal components $g_{\textrm{xx}}$ and $g_{\textrm{yy}}$ of the quantum metric tensor of the Rashba bands at zero and nonzero planar magnetic field. The maps were calculated assuming a Rashba parameter $\alpha_{\textrm{R}} = 0.4\epsilon_{\textrm{F}}/k_{\textrm{F}}$, with $\epsilon_{\textrm{F}}$ and $k_{\textrm{F}}$ the Fermi energy and Fermi wavevector, respectively. (\textbf{C}) Contour plots of the dipole density components $\Lambda_{\textrm{xxx}}$ and $\Lambda_{\textrm{yyy}}$ of the band-normalized quantum metric in the exclusion region between the two Fermi lines. In response to an electric field $E^{\omega}$, the planar magnetic field activates a nonlinear longitudinal current $j^{2\omega}$ driven by the band-normalized quantum metric.} 
\label{fig1}
\end{figure*}

We consider next the effect of a planar magnetic field $B$, which, without loss of generality, we set along the $y$ direction. The Kramers doublet at ${\bf k} \neq 0$ is now split, but a mirror symmetry-protected double degeneracy still occurs on the residual mirror-symmetric line $k_{\textrm{y}} = 0$. The ensuing distortion of the spin-split bands is accompanied by an analogous change of the spin textures, which still preserve a mirror-symmetric arrangement. 
Thus, the only effect of the magnetic field on the quantum metric is a simple shift of its singular point and lines of zeros. However, the consequences for the dipole of the BNQM are completely different. In the distorted annulus between the two Fermi lines of the spin-split bands, the BNQM dipoles lose their quadrupolar profile. As a result, the integrated dipole component $\Lambda_{\textrm{xxx}}$ governing the nonlinear response perpendicular to the magnetic field becomes finite. Its magnitude increases linearly with $B$ for small magnetic fields and diverges at the critical field $B_{\textrm{c}}$ at which the Fermi lines of the two split-spin bands intersect each other on the band degeneracy point \cite{suppScience}. For magnetic fields much larger than $B_{\textrm{c}}$, the nonlinear response becomes again vanishingly small.

The spin-momentum locking is therefore endowed with a quantum metric that only requires broken inversion symmetry and exists in time-reversal symmetric conditions. However, electronic transport associated with the quantum metric becomes possible only upon lifting the time reversal symmetry \cite{Kaplan2022,Lahiri2023}. This can be accomplished by applying a planar magnetic field, which activates a nonlinear in-plane magnetoresistance that can be thus used as a probe of the quantum metric. We call this effect quantum metric magnetoresistance (QMMR). Differently from topological antiferromagnets \cite{Gao2023,Wang2023c}, the QMMR is allowed in materials lacking $\mathcal{P}\mathcal{T}$ symmetry. Importantly, the BNQM dipole component $\Lambda_{\textrm{yyy}}$ regulating the nonlinear conductivity parallel to the applied magnetic field remains overall zero, which implies the absence of a nonlinear magnetoresistance in the direction of $B$. This features makes the QMMR similar to the semiclassical Drude-like bilinear magnetoelectric resistance (BMER) first unveiled at the surfaces of three-dimensional topological insulators \cite{He2018a}, but with one important difference. The BMER of the Rashba spin-split bands is strictly zero for magnetic fields smaller than $B_{\textrm{c}}$ ~\cite{suppScience,Tuvia2023}, thus making the nonlinear in-plane magnetoresistance at small fields completely determined by the BNQM.  

\begin{figure*}[!tb]
\centering
\includegraphics{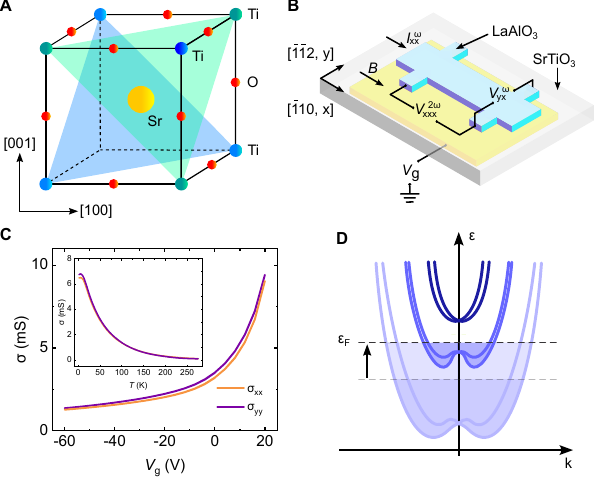}
\caption{\textbf{2D electron gas at the 111-oriented LaAlO$_3$/SrTiO$_3$ interface.} (\textbf{A}) Crystal structure of SrTiO$_3$. The shaded triangles identify the \{111\} planes of Ti atoms. (\textbf{B}) Sketch of the magnetotransport measurements. The transverse and longitudinal first and second harmonic voltages are measured in a Hall bar device while sweeping the magnetic field along the $[\bar{1}10]$ $(x)$ or $[\bar{1}\bar{1}2]$ $(y)$ direction. Two devices with the current path oriented along $x$ and $y$ are measured simultaneously. A variable gate voltage $V_{\textrm{g}}$ is applied at the back of the sample. (\textbf{C}) Gate voltage and temperature (inset) dependence of the sheet conductivity. (\textbf{D}) Schematic representation of the (exaggerated) Rashba-split electronic bands at the 111-oriented LaAlO$_3$/SrTiO$_3$ interface. The arrow shows the direction of band filling upon application of a positive gate voltage.}
\label{fig2}
\end{figure*}

\paragraph*{Nonlinear magnetotransport}
To probe this quantum magnetotransport effect, we consider the two-dimensional electron gas at the 111-oriented LaAlO$_3$/SrTiO$_3$ interface (Fig. 2A). The heterostructures are synthesized by pulsed laser deposition and lithographically patterned into Hall bars oriented along the two principal in-plane crystallographic directions: the $[\bar{1}10]$ $(x)$ and the $[\bar{1}\bar{1}2]$ $(y)$ axis (Fig. 2B) \cite{suppScience}. The sheet conductivity of the electron gas is controlled by an electrostatic field effect in a back-gate geometry that tunes the carrier density and mobility (Fig. 2C) \cite{Caviglia2008,Monteiro2019}. The combination of spin-orbit coupling and orbital degrees of freedom associated with the $t_{\textrm{2g}}$ electrons of Ti atoms leads to three Kramers related pairs of Rashba bands, enabling a gate-induced Lifshitz transition from one- to multi-carrier transport (Fig. 2D) \cite{Rodel2014,Khanna2019,Diez2015}. The Rashba effect endows all three band pairs with the quantum metric and BNQM dipoles discussed above. However, the nonlinear magnetotransport driven by the quantum metric is expected to be always dominated by one pair of bands only. This is because the critical magnetic field $B_{\textrm{c}}$ depends strongly on the band filling and is of the order of a few Teslas when the chemical potential is close to the band bottom of one of the Kramers pairs \cite{suppScience}. Stated differently, at intermediate fields only low-filled Rashba bands yield a sizable QMMR.

\begin{figure*}[!tb]
\centering
\includegraphics{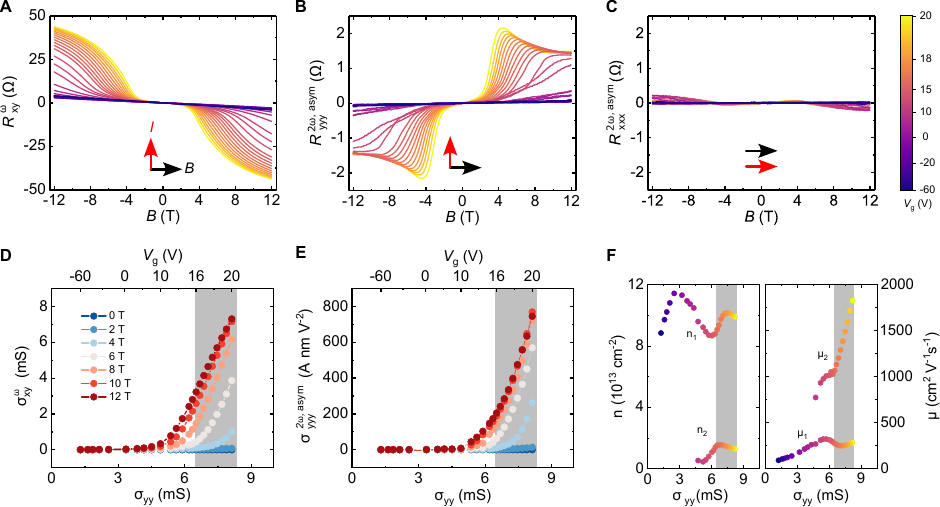}
\caption{\textbf{Linear and nonlinear magnetotransport.} (\textbf{A-C}) First harmonic transverse and second harmonic longitudinal resistances as a function of the in-plane magnetic field $B$ and gate voltage $V_{\textrm{g}}$ at a temperature $T = 3$ K.  The black and red arrows indicate the direction of $B$ and the electric current $I$, respectively. The current is applied along the $\bar{1}\bar{1}2$ $(y)$ crystallographic direction in \textbf{A, B} and along $\bar{1}10$ $(x)$ in \textbf{C}. (\textbf{D}) Transverse linear conductivity calculated from \textbf{A} at different magnetic fields as a function of the zero-field longitudinal linear conductivity. (\textbf{E}) As in \textbf{D}, but for the longitudinal nonlinear conductivity calculated from \textbf{B}. (\textbf{F}) Two-band electron densities $n$ and mobilities $\mu$ at variable $V_{\textrm{g}}$ and $T = 3$ K. The shaded areas define the conductivity regions considered in Fig. 4.}
\label{fig3}
\end{figure*}

Figure \ref{fig3} showcases the electronic response of the 2D electron gas as a function of the planar magnetic field oriented along the $[\bar{1}10]$ direction. At fields well below 3~T, the first harmonic transverse resistance $R^{\omega}_{\textrm{xy}}$ increases linearly with the magnetic field strength, which is compatible with a small out-of-plane misalignment of the field (Fig. 3A). However, $R^{\omega}_{\textrm{xy}}$ increases sharply above a magnetic field threshold modulated by the back gate. This sudden onset can be attributed to the anomalous planar Hall effect appearing in systems without mirror symmetries~\cite{Battilomo2021,Trama2022b}. We note that, although the magnetic field along the $[\bar{1}10]$ direction leaves a residual mirror symmetry on the triangular Ti atoms net, the antiferrodistortive octahedron rotations of the oxygen atoms and the formation of domain patterns at the cubic-to-tetragonal transition of SrTiO$_3$ make our system mirror-free independently of the magnetic field direction. This then leads to the emergence of Berry curvature hotspots when the avoided level crossing at $B_{c}$ enters the Fermi surface annulus~\cite{Lesne2022a}.
The appearance of the anomalous planar Hall effect at fields of a few Teslas thus confirms that the critical magnetic field $ B_{\textrm{c}}$ is experimentally accessible and that the chemical potential lies close to the bottom of one of the Kramers pairs.

At magnetic field strengths $B \geq 2$ T, we also observe a nonlinear and field-antisymmetric in-plane magnetoresistance $R^{2\omega, \, \textrm{asym}}_{\textrm{yyy}}$ (Fig. 3B) that is strongly suppressed when the magnetic field is collinear with the current (Fig. 3C). Sweeping the gate voltage allows us to tune $R^{2\omega, \, \textrm{asym}}_{\textrm{yyy}}$ and explore its dependence on the band filling. As shown in Fig. 3E, the nonlinear magnetoconductivity $\sigma^{2\omega}_{\textrm{yyy}} = J^{2\omega}_{\textrm{y}}/(E^{\omega}_{\textrm{y}})^2$ associated with $R^{2\omega, \, \textrm{asym}}_{\textrm{yyy}}$ vanishes in the low conductivity region and sharply emerges at a threshold $\sigma_{\textrm{yy}} = J^{\omega}_{\textrm{y}}/E^{\omega}_{\textrm{y}} \simeq 6 $~mS, similar to the anomalous planar Hall conductivity $\sigma^{\omega}_{\textrm{xy}} = J^{\omega}_{\textrm{x}}/E^{\omega}_{\textrm{y}}$ (Fig. 3D). Measurements of the ordinary Hall effect reveal that a transition from one to two-carrier transport occurs at this threshold conductivity (Fig. 3F) ~\cite{suppScience,Monteiro2019,Khanna2019}. This finding indicates that the enhancement of the nonlinear response practically coincides with the filling of the second pair of Kramers bands. All together, these observations demonstrate the existence of a nonlinear longitudinal electronic response that fulfils the requirements of the BNQM-driven magnetotransport. We note that the field-symmetric counterpart of $R^{2\omega, \, \textrm{asym}}_{\textrm{yyy}}$ cannot originate from the quantum metric but derives from nonlinear skew scattering and side-jump contributions that exist also in the presence of time-reversal symmetry ~\cite{suppScience,Du2019,Ortix2021}.

\begin{figure*}[!tb]
\centering
\includegraphics{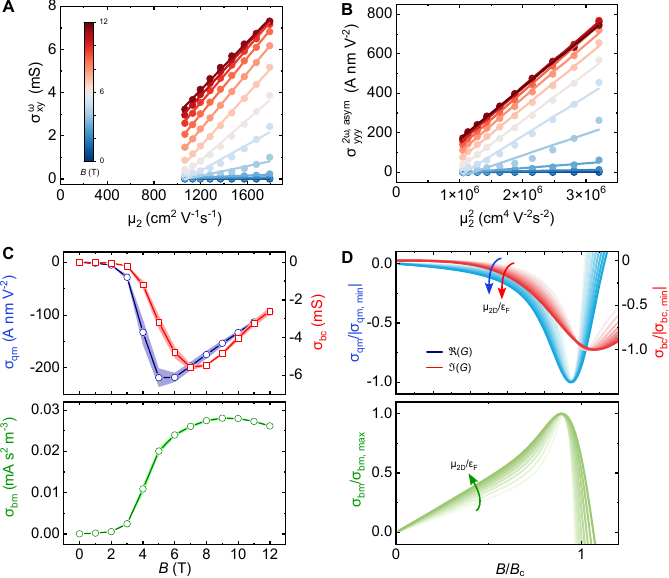}
\caption{\textbf{Electronic transport driven by the full quantum geometric tensor.} (\textbf{A}) Linear transverse conductivity at different magnetic fields as a function of the electronic mobility of the second Kramers pair. The lines are linear fits to the data. (\textbf{B}) Nonlinear longitudinal conductivity as a function of the square of the electronic mobility of the second Kramers pair. (\textbf{C}) Quantum metric ($\sigma_{\textrm{qm}}$), Berry curvature ($\sigma_{\textrm{bc}}$), and bilinear magnetoeletric ($\sigma_{\textrm{bm}}$) conductivities extracted from the fits in \textbf{A, B} as a function of the magnetic field. The shaded areas define the uncertainty of the fits. (\textbf{D}) Magnetic field dependence of the normalized $\sigma_{\textrm{qm}}$, $\sigma_{\textrm{bc}}$, and $\sigma_{\textrm{bm}}$ calculated from $\sigma_{\textrm{qm}}=\frac{e^3}{\hbar}\sum_n \int_{\mathbf{k}} f_n(\mathbf{k}) \Lambda^{(n)}_{xxx}(\mathbf{k})$, $\sigma_{\textrm{bc}}= - \frac{e^2}{\hbar}\sum_n \int_{\mathbf{k}} f_n(\mathbf{k}) \Omega(\mathbf{k})$ and $\sigma_{\textrm{bm}}= -\tau^2\frac{e^3}{\hbar^3}\sum_n\int_{\mathbf{k}} f_n(\mathbf{k}) {\partial^3\varepsilon_n(\mathbf{k})}/{\partial k_x^3}$, with $n$ band index and  $f_n(\mathbf{k})$  the Fermi distribution function \cite{suppScience}. The color intensity codes the ratio $\mu_{\textrm{2D}}/\epsilon_{\textrm{F}} = 0.2-0.6$ between the chemical potential of the 2D electron gas and the Fermi energy $\epsilon_{\textrm{F}}$. The magnetic field is normalized to the critical field $B_{\textrm{c}}$.} 
\label{fig4}
\end{figure*}

\paragraph*{Quantum metric nonlinear magnetoresistance}
Our experimental observations cannot be attributed to thermal effects, nonohmic contacts, capacitive or inductive coupling, or the AC modulation of the gate voltage \cite{suppScience}. Instead, the nonlinear in-plane magnetoresistance is compatible with the QMMR of Rashba spin-split bands and the BMER. We note that, because of hexagonal warping effects in 111-oriented heterostructures, the BMER can be finite not only for magnetic field strengths larger than the critical magnetic  field but also for $B<B_{\textrm{c}}$~\cite{Tuvia2023}. 
The observation of a quantum metric-induced nonlinear transport, therefore, requires to disentangle the QMMR from the BMER. To this aim, we exploit the fact that
the QMMR and BMER contributions to the total nonlinear magnetoresistance can be parsed thanks to their different dependence on the electronic scattering time $\tau$~\cite{Kaplan2022}. The latter can be controlled via the gate tunability of the sheet conductance or by varying temperature. Whereas the QMMR contribution $\sigma_{\textrm{qm}}$ is intrinsic in origin and scales as $\tau^0$, the BMER grows with the square of the scattering time, i.e., as $\sim\tau^2$. Because the mobility is linear in the band-resolved relaxation time, we can equivalently consider the dependence of the nonlinear conductivity on the mobility $\mu_2$ of the second Kramers pair of bands. As shown in Fig. 4B, this scaling analysis allows us to disentangle the two effects in the conductivity region in which the carrier density $n_2$ is constant \cite{Wang2023c}. The nonideal linear fit at intermediate magnetic fields may be attributed to the variation of the spin-orbit coupling and effective mass with the gate voltage, which is not captured by the mobility scaling analysis. We find that both the QMMR and the BMER conductivities change nonmonotonically with the magnetic field (Fig. 4C), in agreement with our theory calculations based on a Rashba two-dimensional electron gas with an additional trigonal warping (Fig. 4D) \cite{suppScience}. The opposite sign for the QMMR and BMER predicted by the theoretical analysis also matches our experimental observation. We can then conclude that the Rashba electron gas at the 111-oriented LaAlO$_3$/SrTiO$_3$ interface features a nonlinear magnetoresistance driven by a quantum metric contribution that is three orders of magnitude larger than that reported for 2D topological antiferromagnets at a similar temperature \cite{Wang2023c}.

Finally, we have performed a similar scaling analysis of the anomalous planar Hall resistance to account for the presence of disorder-mediated effects beside the intrinsic Berry curvature-induced contribution. The linear dependence of the anomalous planar Hall conductivity on the mobility $\mu_2$ (Fig. 4A) demonstrates the existence of a skew scattering contribution $\sigma_{\textrm{sk}}$ that is absent in a rotational symmetric Rashba two-dimensional electron system~\cite{Borunda2007} but is activated by the nonvanishing Berry curvature triple owing to hexagonal warping effects~\cite{Makushko2023}. In addition, the $\tau^0$ term due to the combined action of side-jump and intrinsic Berry curvature varies nonmonotonically as the magnetic field is increased. This behavior is correctly captured by a theory calculation of the Berry curvature-mediated anomalous planar Hall response (Fig. 4D), which indicates that the intrinsic contribution dominates the $\tau$-independent anomalous planar Hall conductivity. Therefore, the intrinsic contributions to the anomalous planar Hall effect and nonlinear in-plane magnetoresistance provide us with direct access to the Berry curvature and the quantum metric of the 111-oriented LaAlO$_3$/SrTiO$_3$ interface and allow us to probe electronic dynamics associated with its full quantum geometric tensor. This unique property marks the difference of 111-oriented LaAlO$_3$/SrTiO$_3$ with respect to $\mathcal{P}\mathcal{T}$-symmetric materials like topological antiferromagnets \cite{Gao2023,Wang2023c}, where the quantum geometric tensor is purely real and accessible only via the nonlinear response.

\paragraph*{Discussion and outlook}
The analysis of the conductivity dependence on the scattering time allowed us to identify the intrinsic contribution of the quantum metric and Berry curvature to the electronic magnetotransport. Whereas the Berry curvature is enabled by the trigonal symmetry of the 111-oriented interface, the quantum metric is rotationally invariant and is not contingent on specific symmetry requirements beside the broken inversion symmetry. This property makes quantum metric effects potentially observable in a vast class of nonmagnetic materials. Moreover, the common origin of the QMMR and BMER suggests that all platforms in which the BMER was previously reported also host the quantum metric. The QMMR is thus not limited to oxide heterostructures~\cite{Ohtomo2004a,Vicente-Arche2021} but is expected for all materials with spin-locked electronic bands such as semiconductors \cite{Nitta1997,Li2014,Ideue2017,Li2021e}, surface states of heavy metals \cite{Ast2007}, magnetic and nonmagnetic interfaces \cite{Li2014,Bihlmayer2022,Sanchez2013}, and even policrystalline samples. Our findings, therefore, do not only demonstrate the existence of a quantum metric associated with the spin-momentum locking but suggest new directions for investigating and functionalizing the quantum geometry. Besides magnetotransport, the quantum metric of the spin-momentum locking is, in principle, accessible to THz photocurrents and spectroscopic tools in time-reversal symmetric conditions ~\cite{Ahn2020,Ma2023a}. From a technological perspective, integrating the spin-momentum locking with magnetic and ferroelectric thin films could lead to the on-demand, reversible, and nonvolatile control of the physical effects emerging from the quantum metric. Moreover, combining spintronics with the quantum metric may reveal unforeseen disorder-free mechanisms to interconvert charge and spin-orbital currents \cite{Li2014,Bihlmayer2022,Feng2024}. This study, therefore, may enable a range of possibilities to harness quantum metric effects in condensed matter and may ultimately inspire technologies based on the geometry of the electronic wavefunctions.

\bibliographystyle{Science}


\section*{ACKNOWLEDGMENTS}
We thank Jean-Marc Triscone, Marc Gabay, Ajit Srivastava, Edouard Lesne, and Graham Kimbell for fruitful discussions. We thank Marco Lopes and Jean-Luc Lorenzoni for technical support. We thank Alberto Morpurgo and Ignacio Gutiérrez for contributing to the sample fabrication. \textbf{Funding:} This work was supported by the Swiss State Secretariat for Education, Research and Innovation (SERI) under contract no. MB22.00071, by the Gordon and Betty Moore Foundation (grant no. 332 GBMF10451 to A.D.C.), by the European Research Council (ERC), and by the Dutch Research Council (NWO) as part of the VIDI (project 016.Vidi.189.061 to A.D.C.). G.S. acknowledges support from the Swiss National Science Foundation (grant no. PZ00P2\_223542). A.D.C, M.T.M and M.C. acknowledge support from the EU Horizon 2020 research and innovation program under Grant Agreement No. 964398 (SUPERGATE). M.T.M. and C.O. acknowledge support from the MAECI project "ULTRAQMAT" and the PNRR MUR project PE0000023-NQSTI (TOPQIN). M.C. acknowledges financial support from PNRR MUR project PE0000023-NQSTI. \textbf{Author contributions:} G.S. fabricated the oxide heterostructures and the transport devices used in this study. G.S. and K.D. performed the measurements and analyzed the data with the support of S.G.. M.T.M., M.C., and C.O. developed the theoretical model. A.D.C. supervised the experimental work, and C.O. supervised the theoretical work. All authors contributed to writing the manuscript. \textbf{Competing interests:} The authors declare no competing financial interests.  \textbf{Data and materials availability:} The data and code that support the findings of this study are available from Zenodo \cite{zenodo_Science2024}.

\nocite{Bures1969,Hikami1980,Maekawa1981,Xu2024z}

\section*{SUPPLEMENTARY MATERIALS}
Materials and Methods\\
Supplementary Text\\
Figs. S1 to S13\\
References \textit{(56-59)}



\clearpage

\noindent\LARGE{\textbf{Supplementary Materials}} \\
\\
\noindent\normalsize{\textbf{The PDF file includes:}} \\
\\

\noindent Materials and Methods\\
\noindent Supplementary Text\\
\noindent Figs. S1 to S13\\
\noindent References \textit{(56-59)}\\

\clearpage

\noindent\textbf{Materials and Methods} \\

\noindent\underline{Sample growth and device characterization} \\
The 2D electron system at the 111-oriented LaAlO$_3$/SrTiO$_3$ interface was prepared by depositing an epitaxial LaAlO$_3$ layer on commercially available SrTiO$_3$ substrates. The substrate surface was processed with a sequence of di-H$_2$O bleaching and air annealing steps to obtain a terraced structure. 9 unit cells of LaAlO$_3$ were grown by pulsed laser deposition using a KrF excimer laser (wavelength: 248 nm, fluence: 1.1 mJ cm$^{-2}$, repetition rate: 1 Hz) in an O$_2$ pressure of 8$\cdot10^{-5}$ Torr and at a temperature of 700 °C. The layer-by-layer growth was monitored by observing the intensity oscillations of the high-energy electron diffraction. After the growth, the samples were annealed at 500 °C for an hour in a 300 mbar O$_2$ atmosphere and subsequently cooled down to room temperature at a nominal rate of 20 °C/min. 

Hall bar devices were fabricated by a combination of optical lithography and Ar ion milling (Fig. S1). The milling time was calibrated to stop the etching at the LaAlO$_3$/SrTiO$_3$ interface. Ti/Au contact pads were deposited by electron beam evaporation. The Hall bars were patterned along the $[\bar{1}10]$ and $[\bar{1}\bar{1}2]$ crystallographic directions of the substrate. The Hall bars have a width $W = 20$ \textmu m and a distance between the longitudinal probes $L = 60$ \textmu m. 
\newline

\noindent\underline{Magnetotransport measurements} \\
Devices along the orthogonal $[\bar{1}10]$ and $[\bar{1}\bar{1}2]$ 
directions were contacted with Al wire bonds. Electrically- and thermally-conductive Ag paint was used to anchor the samples to the chip carrier and apply a back-gate voltage. Four-probe magnetotransport measurements were performed in a liquid He$^4$ Oxford Teslatron cryostat with base temperature of 1.7 K equipped with a superconducting magnet capable of generating magnetic fields up to 12 T. Nonlinear transport measurements were performed with the magnetic field applied in the sample plane along one of the two crystalline directions. Measurements of the ordinary Hall effect were performed with the magnetic field applied in the direction normal to the sample surface. The magnetic field was typically swept at a rate of 0.15 T/min. Commercial lock-in amplifiers (Zurich Instruments MFLI and Stanford Research 830) were used to apply an alternate current $I = 50$ \textmu A at a frequency $f = \frac{\omega}{2\pi} = 17.7$ Hz and simultaneously detect the longitudinal and transverse harmonic voltages in both Hall bars. The amplitude of the electric current was controlled by setting the lock-in output voltage at a few Volts and by connecting a 100 kOhm resistor in series to the Hall bars. The actual current through the device was measured by terminating the circuit on the 50-Ohm current input of the lock-in amplifier. The DC back gate voltage was applied via a Keithley 2400 Sourcemeter. Prior to the measurements, an initial gate-forming procedure was followed to avoid hysteretic variations of the device resistance. \newline

\noindent\underline{Analysis of the magnetotransport measurements} \\
The measured signals were symmetrized and antisymmetrized with respect to the magnetic field $B$ to isolate distinct effects according to
\begin{align}
\chi^{\textrm{sym}} = \frac{\chi(B) + \chi(-B)}{2}\\
\chi^{\textrm{asym}} = \frac{\chi(B) - \chi(-B)}{2}
\end{align}
The first harmonic longitudinal resistance has only a symmetric contribution, therefore $R^{\omega, \, \textrm{asym}}_{ii} \equiv R^{\omega}_{ii}$, with $i = x, \,y$. The first harmonic transverse resistance includes both symmetric and antisymmetric contributions. Since the symmetric part originates only from the contact misalignment and the anisotropic magnetoresistance, we neglect it and use the simplified notation $R^{\omega, \, \textrm{asym}}_{ji} = R^{\omega}_{ji}$. The second harmonic longitudinal resistance also includes symmetric and antisymmetric components. Since the nonlinear electronic transport driven by the quantum metric requires broken time reversal, only the antisymmetric part is considered in the main text. The symmetric component is discussed in the supplementary material.

The first and second harmonic resistances were calculated from the measured voltages as $R^{\omega}_{ji} = \frac{V^{\omega}_{ji}}{I_{i}^{\omega}}$ and $R^{2\omega}_{jii} = \frac{V^{2\omega}_{jii}}{I_{i}^{\omega}}$, respectively, with $i, j = x, \,y$. $i$ is the direction of the injected current, and $j$ is the direction of the measured voltage. Longitudinal (transverse) signals correspond to $i = j$ ($i \neq j$). The sheet conductivity was calculated from the first harmonic longitudinal resistance as $\sigma_{ii} = \frac{L}{WR^{\omega}_{ii}}$. The nonlinear conductivity $\sigma^{2\omega}_{iii}$ was defined as $J^{2\omega}_{i} = \sigma^{2\omega}_{iii}(E_{i}^{\omega})^2 = \sigma_{ii}E_{i}^{2\omega}$, where $J$ and $E$ are the current density and electric field, respectively. The longitudinal nonlinear conductivity was therefore obtained from the relation $\sigma^{2\omega}_{iii} = \frac{R^{2\omega}_{iii}}{(R^{\omega}_{ii})^3I_{i}^{\omega}}\frac{L^2}{W}$.

The carrier density $n$ and electron mobility $\mu$ were obtained by fitting the ordinary Hall resistance measured with an out-of-plane magnetic field $B$ to the two-band model:
\begin{equation}
 R^{\omega}_{ji} = \frac{(\sigma_1\mu_1 + \sigma_2\mu_2) + (\sigma_1\mu_2 + \sigma_2\mu_1)\mu_1\mu_2B^2}{(\sigma_1+\sigma_2)^2 + (\sigma_1\mu_2 + \sigma_2\mu_1)^2B^2}B.
\end{equation}
The conductivities of the two bands were constrained by setting $\sigma_1 + \sigma_2 = \sigma_{ii}.$ The carrier density was calculated as $n_{1,2} = \frac{\sigma_{1,2}}{e\mu_{1,2}}$, where $e$ is the electron charge.
\newline

\clearpage
\noindent\textbf{Supplementary Text}
\newline

\noindent\underline{1 Quantum geometry of spin-locked electronic bands}
\\

\noindent\textit{1.1 The quantum geometric tensor of a Bloch wave}
\\
\noindent
The quantum geometric tensor (QGT) characterizing the cell-periodic part of a Bloch wave $|u_n(\bk)\rangle$ that satisfies the Schr\"odinger equation ${\mathcal H}(\bk) |u_n(\bk)\rangle =\varepsilon_n(\bk)|u_n(\bk)\rangle$, with ${\mathcal H}(\bk)$ the Bloch Hamiltonian and $n$ a band index, can be introduced by recalling that treating the $d$-dimensional crystalline momentum $\bk$ as a parameter, the distance between two quantum states in parametric space is defined as \textit{(15,56)};
\begin{equation}
    D^2_{n;12}=1-|\langle u_n(\bk_1) |u_n(\bk_2) \rangle |^2\;.
\end{equation}
The infinitesimal distance in the Hilbert space can then be written as 
\begin{equation}
D^2_{n;\bk,\bk+d\bk}=\sum_{a,b=1}^d g_{n,ab}\,dk_a\, dk_b,
\end{equation}
which defines a quantum metric tensor $g_{n,ij}(\bk)$ (12), 
that reads as:
\begin{eqnarray}
\label{eq:qm}
    g_{n,ab} = \text{Re} \langle \partial_{a}u_n(\bk) | \hat{Q}_n(\bk) |\partial_{b} u_n(\bk)\rangle \;,
\end{eqnarray}
where $\hat{Q}(\bk)=\hat{1}-\hat{P}( \bk)$, is the complement of the eigenstate projector $\hat{P}(\bk) =|u_n(\bk)\rangle \langle u_n(\bk)|$
and $\partial_j\equiv \partial/\partial k_j$.
The quantum metric is thus the real (symmetric) part of the QGT $T_{n,ab}$:
\begin{eqnarray}
    T_{n,ab}(\bk) &=&  \langle \partial_{a}u_n(\bk) | \hat{Q}_n(\bk) |\partial_{b} u_n(\bk)\rangle \\
    &=&\sum_{m\neq n} \mathcal{A}^a_{nm}\mathcal{A}^b_{mn}\;,
\end{eqnarray}
where  $\mathcal{A}^a_{nm} = i \langle u_n | \partial_a u_m\rangle$ is the interband Berry connection. 
Note that the imaginary part of the QGT can be simply related to the Berry curvature $\Omega_{n,ab}$. In particular, one has that the quantum geometric tensor
$T_{n,ab} = g_{n,ab}-\frac{i}{2}\Omega_{n,ab}$.

Another important quantity, which is expressed in terms of the interband Berry connection  $\mathcal{A}^a_{nm}$, is the Berry connection polarizability 
\begin{eqnarray}
\label{eq:Gab1}
    G_n^{ab} &=& 2 \text{Re} \sum_{m\neq n}\frac{\mathcal{A}^a_{nm}\mathcal{A}^b_{mn}}{\varepsilon_n-\varepsilon_m}\; ,
\end{eqnarray}
which is also called band-energy normalized quantum metric (BNQM).
\newline

\noindent\textit{1.2 Nonlinear magnetotransport in a Rashba two-dimensional electron system}
\\
\noindent
The nonlinear longitudinal magnetotransport can be generally written as \textit{(11)}
\begin{eqnarray}
    J_a &=& \sigma_{aaa} E_a^2 \quad (a=x,y)\\
    \sigma_{aaa} &=& \sigma^{(\rm{qm})}_{aaa} +  \sigma^{(\rm{bm})}_{aaa} + \sigma^{(\rm{sj})}_{aaa} + \sigma^{(\rm{sk})}_{aaa}
\end{eqnarray}
The last two terms in the equation above correspond to the nonlinear side-jump and skew scattering contributions \textit{(8)}
that exist in time-reversal symmetric conditions, i.e., at $B \equiv 0$, and have been probed at the trigonal surfaces of bismuth thin films \textit{(9)}.
These terms are expected to be symmetric in the Zeeman coupling $B$. On the contrary, the first two terms must vanish in the presence of time-reversal symmetry and are therefore, antisymmetric, in $B$. In particular, $\sigma^{(\rm{qm})}_{aaa}$ is the intrinsic contribution $\propto \tau^0$, with $\tau$ the relaxation time, that is generated by the quantum metric. The nonlinear Drude weight given by $\sigma^{(\rm{bm})}_{aaa}$ scales as $\tau^2$ and is responsible for the bilinear magnetoelectric resistance. These two terms explicitly read: 
\begin{eqnarray}
\label{eq:sigmaQM}
      \sigma^{(\rm{qm})}_{aaa}&=&=\frac{e^3}{\hbar}\sum_{n=\pm} \int_\bk  f_n(\bk)\frac32\partial_{k_a} G_n^{aa} \equiv \frac{e^3}{\hbar}\sum_n \int_{\bk} f_n(\bk) \Lambda^{(n)}_{aaa}(\bk) \;,\\
\label{eq:sigmaBM}
       \sigma^{(\rm{bm})}_{aaa}&=& -\tau^2\frac{e^3}{\hbar^3}\sum_n\int_{\bk} f_n(\bk) \frac{\partial^3\varepsilon_n(\bk)}{\partial k_a^3} \;.
\end{eqnarray}
In the equations above $f_n(\bk)$ is the Fermi distribution function, $\int_\bk\equiv\int \frac{d^2k}{(2\pi)^2}$, $G_n^{aa}$ is the BNQM (see Eq. \ref{eq:Gab1}) and the kernel $\Lambda_{aaa}^{(n)}$ introduced in Eq. \ref{eq:sigmaQM} is proportional to the dipole of the BNQM $\partial_{k_a}G^{aa}_n$.  

We consider then the effective Hamiltonian for a conventional two-dimensional electron system with spin-momentum locking. This is given by the Rashba Hamiltonian, which, when augmented with a Zeeman coupling due to a planar magnetic field, can be written as 
\begin{eqnarray}
    \label{eq:Ham0}
H(\bk)&=& \frac{\hbar^2 k^2}{2 m^{\star}} +\alpha_R(k_x\sigma_y - k_y\sigma_x) + \mathbf{B}\cdot\boldsymbol{\sigma} \\
&=&h_0(\bk)\sigma_0 + \bh(\bk)\cdot\boldsymbol{\sigma}\;,\qquad \quad {\text{ with  }}\bh=\{-\alpha_R k_y+B_x, \;\alpha_R k_x + B_y,\;0\}\;. 
\label{eq:hvec}
\end{eqnarray}
In the equation above $\sigma_i\;(i=x,y,z)$ are the Pauli matrices, $\sigma_0$ is the identity matrix, $m^{\star}$ is the effective electron mass,  $\alpha_R$  is the Rashba spin-orbit coupling strength and $B$ is the Zeeman energy. 
Note that the Hamiltonian for the surface states of topological insulators can be recovered by neglecting the identity term $\propto k^2$. 
The Hamiltonian in Eq. \ref{eq:Ham0} can be conveniently written in terms of the vector $\bh(\bk)$ such that the spin-orbit coupled bands can be expressed in terms of this Hamiltonian vector  as $E_\pm(\bk)=h_0(\bk)\pm |\bh (\bk)| $.

The quantum metric and Berry curvature can be expressed in terms of $\hat{\bh}=\bh/h$:
\begin{eqnarray}
 \label{eq:qm0}
    g^\pm_{ab} &=& \frac{1}{4} \partial_{k_a}\hat{\bh}\cdot\partial_{k_b}\hat{\bh} \\ 
    \label{eq:bc0}
    \Omega_\pm&=&\mp \frac{1}{2}\hat{\bh}\cdot (\partial_{k_x}\hat{\bh}\times \partial_{k_y}\hat{\bh})\;,
\end{eqnarray}
where $(\pm)$ are the band indices, and $a$ and $b$ are Cartesian coordinates ($x,y$). For the simple Rashba Hamiltonian in Eq.~\ref{eq:Ham0}  the Berry curvature is zero, while for the quantum metric we have
\begin{eqnarray}
   g_{xx} &=&\frac{\alpha_R^2(\alpha_R k_y -B_x)^2}{4[(\alpha_R k_y -B_x)^2+(\alpha_R k_x+B_y)^2]^2} \\
    g_{xy} &=&\frac{\alpha_R^2 (\alpha_R k_x +B_y)(-\alpha_R k_y +B_x)}{4[(\alpha_R k_y -B_x)^2+(\alpha_R k_x+B_y)^2]^2} \\
    g_{yy} &=&\frac{\alpha_R^2(\alpha_R k_x +B_y)^2}{4[(\alpha_R k_y -B_x)^2+(\alpha_R k_x+B_y)^2]^2}\;.
\end{eqnarray}
Analytical expressions of the BNQM and $\Lambda^{(\pm)}_{aaa}$ follow easily. Notice that, while the quantum metric is the same for the two bands, the BNQM $G_{\pm}^{ab}$ (and hence $\Lambda^{(\pm)}$) for the two bands are opposite in sign and equal in amplitude. This is because for a two band system the BNQM is simply related to the quantum metric by  $G_\pm^{ab} = \pm g_{ab}/{2h}$.  

At zero temperature, the contribution to the nonlinear conductivity $\sigma^{(\rm{qm})}_{aaa}$ is given by the integral of $\Lambda^{(-)}_{aaa}$ in the region between the two Fermi lines, given by the condition $E_\pm(\bk)=\mu$. 
We have 
\begin{eqnarray}
\label{eq:Lxxx}
    \Lambda^{(\pm)}_{xxx} (\bk)&=&\mp \frac{15\alpha_R^3 (\alpha_R k_x +B \sin\theta)(\alpha_R k_y -B \cos\theta)^2}{16[(\alpha_R k_x +B \sin\theta)^2+(\alpha_R k_y -B \cos\theta)^2]^{7/2}} \\
\label{eq:Lyyy}
     \Lambda^{(\pm)}_{yyy} (\bk)&=&\mp \frac{15\alpha_R^3 (\alpha_R k_x +B \sin\theta)^2(\alpha_R k_y -B \cos\theta)}{16[(\alpha_R k_x +B \sin\theta)^2+(\alpha_R k_y -B \cos\theta)^2]^{7/2}} \;,
\end{eqnarray}
where $\theta$ is the angle formed by the planar magnetic field $\bf{B}$ with the $x$ axis. 

For small magnetic fields, it is possible to obtain a closed analytical expression for the quantum metric nonlinear magnetoresistance. 
We first notice that in the presence of the planar Zeeman coupling, it is convenient to introduce the shifted momenta
$(p_x=k_x+(B/\alpha_R) \sin\theta,p_y=k_y-(B/\alpha_R) \cos\theta)$. The spin-momentum locked energy bands then read $E_\pm=h_0(\mathbf{p})\pm |\alpha_R|\sqrt{p_x^2+p_y^2}$  
and cross each other at the shifted origin $p_x=p_y=0$. 
Due to the rotational symmetry of the model Hamiltonian at $B=0$, we can restrict to a single direction $a=x$. We then have 
\begin{eqnarray}
    \Lambda^{(\pm)}_{xxx}(\mathbf{p})&=& \mp \frac{15\alpha_R^6 p_x p_y^2}{16[(\alpha_R p_x)^2+(\alpha_R p_y)^2]^{7/2}} 
    = \mp \frac{15  \cos\phi \sin^2\phi}{ 16|\alpha_R|p^4}\;,
\end{eqnarray}
where in the last expression we used polar coordinates $(p,\phi)$.
Rewriting Eq. \ref{eq:sigmaQM} in terms of $(p,\phi)$, the quantum metric contribution to nonlinear conductivity reads:
\begin{eqnarray}
\sigma_{xxx}^{(\rm{qm})}&=&\frac{e^3}{\hbar} \sum_{n=-,+}\int \frac{dp_x dp_y}{(2\pi)^2} f_n(\mathbf{p})  \Lambda_{xxx}^{(n)}(\mathbf{p}) \nonumber \\
&=&
\frac{e^3}{\hbar} \int \frac{dp_x dp_y}{(2\pi)^2} \Lambda_{xxx}^{(-)}(\mathbf{p})
 (f_-(\mathbf{p})-f_+(\mathbf{p})) \;,
 \end{eqnarray}
where the Fermi-Dirac distribution functions are evaluated, for simplicity, at zero temperature. We next consider a chemical potential $\mu$ such that both spin-orbit coupled bands are occupied.  
Their energies, using polar coordinates, 
are $E_\pm(p,\phi)=\frac{\hbar^2}{2 m}(p^2+\frac{B^2}{\alpha_R^2}+2 p \frac{B}{\alpha_R}\sin(\phi-\theta))\pm p |\alpha_R|$, and the corresponding Fermi lines ($E_\pm = \mu$) can be determined analytically as
\begin{eqnarray}
\label{eq:FLsol1}
    E_-=\mu &\Rightarrow& \qquad 
    p_\pm^{(-)}(\phi)=\frac{1}{2}\left[2\frac{B}{\alpha_R} \sin(\theta-\phi) + |\alpha_R| \right. \nonumber \\
    &\pm&\left. \sqrt{(2\frac{B}{\alpha_R} \sin(\theta-\phi)+|\alpha_R|)^2-4(\frac{B^2}{\alpha_R^2}-\mu)}\right]\;, \\ \label{eq:FLsol2}
    E_+=\mu &\Rightarrow& \qquad
     p_\pm^{(+)}(\phi)=\frac{1}{2}\left[2\frac{B}{\alpha_R} \sin(\theta-\phi)-|\alpha_R| \right. \nonumber \\  &\pm& \left. \sqrt{(2\frac{B}{\alpha_R} \sin(\theta-\phi)-|\alpha_R|)^2-4(\frac{B^2}{\alpha_R^2}-\mu)} \right]\;,
\end{eqnarray}
where, to simplify the notation, we have written energies in units of a reference energy $E_0=(\hbar k_F^0)^2/2m^*$ and wave vectors in units  of $k_F^0$. 

We next introduce the critical magnetic field $B_{\textrm{c}}=\sqrt{\alpha_R^2\mu}$ at which the mirror symmetry protected crossing between the two spin-orbit bands occurs at the Fermi level $E=\mu$. 
For $B<B_{\textrm{c}}$ the two Fermi lines wind around the origin 
of the shifted momenta $p$ space (see  Fig. S2).  
In this regime we have 
\begin{eqnarray}
\sigma_{xxx}^{(\rm{qm})}(B<B_{\textrm{c}})&=&\frac{e^3}{\hbar}  \frac{1}{(2\pi)^2}\int_0^{2\pi} d\phi \int_{p_+^{(+)}(\phi)}^{p_+^{(-)}(\phi)} d\,p  \frac{15  \cos\phi \sin^2\phi}{16 |\alpha_R| p^3} \nonumber \\
&=&\frac{e^3}{\hbar}  \frac{15}{16|\alpha|} \frac{1}{(2\pi)^2}\int_0^{2\pi} d\phi     \cos\phi \sin^2\phi \frac12\left[\frac1{(p_+^{(+)}(\phi))^2}-\frac1{(p_+^{(-)}(\phi))^2}\right]
\nonumber \\
&=&\frac{e^3}{\hbar}  \frac{15}{16|\alpha|} \frac{1}{(2\pi)^2}\int_0^{2\pi} d\phi     \cos\phi \sin^2\phi \left[g_1(\phi)+g_2(\phi)\right] \;,
\end{eqnarray}
where in the last line we separate the terms coming from the integration in $p$ according to their behavior for $\phi \to \phi+\pi$. Indeed, we have $g_1(\phi+\pi)=-g_1(\phi)$ while $g_2(\phi+\pi)=g_2(\phi)$, so only the integral containing $g_1$ will give a non-zero contribution. Using the fact that   $ g_1(\phi) =- 4(B/\alpha_R) |\alpha_R| \sin(\theta-\phi)/(B^2/\alpha_R^2-\mu)^2$, we obtain the analytical result:
\begin{eqnarray}
\sigma_{xxx}^{(\rm{qm})}(B<B_{\textrm{c}})&=&
-\frac{e^3}{\hbar}   \frac{15}{128\pi}  \frac{\frac B{\alpha_R} \sin\theta}{[(\frac B{\alpha_R})^2-\mu]^2}
\label{eq:QMres}
\end{eqnarray}
Using a similar arguments it can be shown that the nonlinear Drude weight $\sigma_{xxx}^{(\rm{bm})}$ is zero for $B<B_{\textrm{c}}$ \textit{(34)}.
Indeed, rewriting the needed expressions in terms of the shifted momenta $\mathbf{p}$, we have
\begin{eqnarray}
\label{eq:SCxxx}
\frac{\partial^3\varepsilon_\pm(\bk)}{\partial k_x^3}
&=&\mp \frac{3\alpha_R^3 (\alpha_R k_x +B \sin\theta)(\alpha_R k_y -B \cos\theta)^2}{[(\alpha_R k_x +B \sin\theta)^2+(\alpha_R k_y -B \cos\theta)^2]^{5/2}} \nonumber \\
   &=&\mp  \frac{3|\alpha_R| p_x p_y^2}{[ p_x^2+ p_y^2]^{5/2}} = \mp \frac{3|\alpha_R|  \cos\phi \sin^2\phi}{ p^2} 
\end{eqnarray}
and hence the bilinear magnetoelectric resistance contribution from Eq. \ref{eq:sigmaBM} reads
\begin{eqnarray}
    \sigma_{xxx}^{(\rm{bm})}&=&\tau^2\frac{e^3}{\hbar^3} \frac{1}{(2\pi)^2}\int d\phi d\,p   \frac{3|\alpha_R|  \cos\phi \sin^2\phi}{ p}(f_+(\mathbf{p})-f_-(\mathbf{p})) \\
\sigma_{xxx}^{(\rm{bm})}(B<B_{\textrm{c}})&=&\tau^2\frac{e^3}{\hbar^3} \frac{3|\alpha_R|}{(2\pi)^2}\int_0^{2\pi} d\phi   \;  \cos\phi \sin^2\phi\;\ln{\left(\frac{p_+^{(-)}(\phi)}{p_+^{(+)}(\phi)}\right)} =0
\end{eqnarray}
The last result is easily obtained by noticing that the ratio $R(\phi)=p_+^{(-)}(\phi)/p_+^{(+)}(\phi)$ is symmetric for $\phi\to\phi+\pi$, 
 thus the integrand  $\ln(R(\phi))\cos\phi \sin^2\phi$ is antisymmetric for $\phi \to \phi +\pi$, resulting in a zero contribution when integrated in $\phi\in[0,2\pi]$ \textit{(34)}.
 We stress that only for $B<B_{\textrm{c}}$ the two Fermi lines wind around the origin of shifted momenta plane (see  Fig. S2), so the same argument cannot be used for $B>B_{\textrm{c}}$.
In conclusion, we find that for $B<B_{\textrm{c}}$ the nonlinear magnetoresistance of a Rashba gas is completely determined by the quantum metric contribution. 

Note that the expression in Eq. \ref{eq:QMres} implies that at  $B=B_{\textrm{c}}$ the conductivity $\sigma_{xxx}^{(\rm{qm})}$ diverges. Such a divergence will be regularized adding to the model in Eq. $\ref{eq:Ham0}$ a small out-of-plane magnetic field $B_z$ or a trigonal warping.
At the same time, the nonlinear Drude weight will acquire a finite component. 

For $B>B_{\textrm{c}}$, as shown in Fig. S2 the integration does not run over all the the angles $\phi\in[0,2\pi]$, so it is not possible to use the same symmetry argument and we resort to numerical results. 
For $B>B_{\textrm{c}}$ we find that $\sigma_{xxx}^{(\rm{qm})}$ changes sign and decrease in amplitude while increasing the field $B_y$.
\\

\noindent\textit{1.3 Effects of trigonal warping}
\\
The model in Eq. \ref{eq:Ham0} does not capture crystalline anisotropy effects, hence we include in our analysis the first symmetry allowed term 
\begin{equation}
\label{eq:HamW}
    H_w(\bk)=\frac\lambda2(k_+^3+k_-^3)\sigma_z
\end{equation}
with $k_\pm=k_x\pm i\, k_y$. The hamiltonian vector in Eq. $\ref{eq:hvec}$ is then modified in
\begin{equation}
\label{eq:hvecW}
    \bh(\bk)= \{-\alpha_{y} k_y +B \cos\theta, \;\alpha_{x}k_x +B \sin\theta, \;\frac{\lambda}{2}(k_x^3-3 k_x k_y^2)\}\;.
\end{equation}
\\
The warping Hamiltonian in Eq. \ref{eq:HamW}, being proportional to  the Pauli matrix $\sigma_z$, will produce an out-of-plane spin texture and a nonzero Berry curvature, 
which is responsible for the anomalous planar Hall effect \textit{(6,40)}.

As mentioned above, the warping term in Eq. \ref{eq:HamW} will also produce  a nonzero contribution to the semiclassical part $\sigma_{xxx}^{(\rm{bm})}$ of the nonlinear conductivity for $B<B_{\textrm{c}}$. In Fig. S3 we show the behavior of $\sigma_{xxx}^{(\rm{qm})}$ and $\sigma_{xxx}^{(\rm{bm})}$ for a representative set of parameters and for four different values of the warping parameter $\lambda$. In  Fig. S3 we also draw (black line) the analytical result in Eq. \ref{eq:QMres} obtained for $\lambda=0$ and $B<B_{\textrm{c}}$. 
\newline

\noindent\textit{1.4 Symmetries and material properties considerations}
\begin{itemize}
\item The quantum metric associated with the spin-momentum locking in Eq. \ref{eq:Ham0} inherits the requirements of the latter, i.e., the broken inversion symmetry, which is the sole necessary condition for a nonzero quantum metric. This implies that the quantum metric does not depend on the specific crystalline anisotropy terms of the material in question. The model in Eq. \ref{eq:Ham0} is indeed rotationally invariant in the absence of planar magnetic fields. This implies the presence of a nonzero quantum metric even in polycrystalline materials with broken inversion symmetry, e.g., evaporated or sputtered multilayered samples. The quantum metric driven by the spin-momentum locking is, therefore, not limited by neither symmetry constraints (apart the broken inversion symmetry) nor by the topology of the material. Detecting the quantum metric in transport experiments, e.g., via the QMMR, requires, however, breaking the time-reversal symmetry. In the specific case considered here, namely, LaAlO$_3$/SrTiO$_3$ interfaces, the inversion symmetry is broken by the interface, and the time reversal symmetry is removed by the magnetic field applied orthogonal to the electric current. The combined PT-symmetry is then also broken.

\item LaAlO$_3$/SrTiO$_3$ features rich electronic and lattice physics and complex structural transitions, but none of these is in general necessary to the existence and detection of the quantum metric associated with the spin-momentum locking. SrTiO$_3$ undergoes a cubic-to-tetragonal transition at about 105 K. When we consider the 111 interface, this transition breaks the three-fold rotational symmetry and only leaves a mirror line. Below 70 K, SrTiO$_3$ experiences an additional tetragonal-to-locally-triclinic distortion. Finally, the onset of a suppressed ferroelectric transition below 40-50 K triggers strong polar quantum fluctuations associated with the displacement of the Ti ions with respect to the oxygen octahedra. These rich structural transitions are crucial for the appearance of the Berry curvature dipole and the associated nonlinear planar Hall effect \textit{(6,40)}. They are instead unnecessary to the existence of the quantum metric driven by the spin-momentum locking.
\newline
\end{itemize}

\noindent\underline{2 Anomalous planar Hall effect}\\

\noindent Figure S4 shows the anomalous planar Hall effect measured while sweeping the magnetic field in the direction parallel or orthogonal to the electric current. The appearance of the planar Hall effect beyond a threshold magnetic field indicates the presence of activated Berry curvature hotspots \textit{(6)}. The signal has similar shape and amplitude independently of the relative orientation of the magnetic field and electric current. This finding and the observation that the planar Hall effect does not vanish when the magnetic field is oriented along the $\bar{1}10$ $(x)$ crystallographic direction indicate the absence of the mirror symmetry $\mathcal{M}_{\bar{1}10}$. This symmetry is possibly broken by the antiferrodistortive rotation of the oxygen octahedra and/or the presence of ferroelastic domains below the cubic-to-tetragonal distortion occurring in SrTiO$_3$ at 105 K. 
\newline

\noindent\underline{3 Ordinary Hall effect} \\

\noindent Figure S5 shows the ordinary Hall effect measured while sweeping the magnetic field in the direction normal to the sample surface. The nonlinearity of the Hall resistance with respect to the magnetic field indicates the presence of two bands at high gate voltage and/or low temperature. We use the two-band model (Methods) to determine the carrier density and mobility of each band as shown in Fig. 3 of the main text. We find that the carrier density and mobility along the $[\bar{1}10]$ and $[\bar{1}\bar{1}2]$ crystallographic directions differ by less than  10\% and show the same dependence on temperature and gate voltage. This suggests a weak dependence of the transport properties on the crystallographic direction in the experimentally accessible range of chemical potential. This observation matches with the phenomenology of the measurements with a planar magnetic field.
\newline 

\noindent\underline{4 Linear magnetoresistance}\\

\noindent Figure S6 shows the longitudinal magnetoresistance, i.e., the variation of the first harmonic longitudinal resistance with the magnetic field at different gate voltages. The magnetoresistance changes with the direction of the magnetic field relative to the sample plane. 

When the field is oriented out of the plane, the magnetoresistance is determined by the combination of the ordinary Lorentz magnetoresistance and weak (anti)localization effects \textit{(57,58)}.
The analysis of this magnetoresistance allows us to extract the strength of the Rashba coupling, as discussed in Supplementary Section 5. The details of the analysis can be found in Ref. \textit{(6)}.

When the magnetic field is oriented in the sample plane, the magnetoresistance is mostly negative. As described in Ref. \textit{(6,39)},
the decrease of the resistance at high magnetic field is caused by a field-induced Lifshitz transition that suppresses interband scattering and enhances the sample conductance. 

The magnetoresistances measured along orthogonal crystallographic directions are similar. \newline 

\noindent\underline{5 Energy scales of the 2D electron gas at the LaAlO$_3$/SrTiO$_3$ interface} \\

\noindent As discussed in the main text, the nonlinear transport is maximized when the degeneracy point of the two Rashba-split bands touches the Fermi level $E_{\textrm{F}}$. The position and the field-induced shift of this degeneracy point depend on the intrinsic energy scales of the Rashba bands, namely, $E_{\textrm{F}}$, the strength of the Rashba coupling, and the Zeeman interaction. Here, we elaborate on this aspect.

Figure S7 shows the Fermi energy of the two lowest Rashba pairs that form the low-energy electronic states of the 2D electron gas at the LaAlO$_3$/SrTiO$_3$ interface. The Fermi energy was calculated from the carrier densities extracted from the ordinary Hall effect by assuming a free electron model with an effective mass of $m^* = 3.3m_0$, where $m_0$ is the electron mass \textit{(37)}.
This approach provides an estimate of the order of magnitude of $E_{\textrm{F}}$. We find a Fermi energy in the order of several tens (a few tens) of meV for the first (second) Rashba doublet. As discussed in the main text, the first doublet is always populated whereas the second doublet only appears above a conductivity of approximately 5-6 mS independently of whether the gate voltage or temperature is modified. 


We consider next the strength of the Rashba coupling, which can be extracted from the weak (anti)localization correction to the magnetoresistance in Fig. S6. Note that this estimate is approximate because the model commonly used to describe weak (anti)localization effects does not take into account the multiband nature of the electronic transport in our system \textit{(57,58)}. With this limitation, we estimate a Rashba parameter $\alpha_{\textrm{R}}$ of the order of 0.2-0.8 meV$\cdot$nm, which corresponds to a spin-orbit splitting $E_{\textrm{SO}} = 2\alpha_{\textrm{R}}k_{\textrm{F}}$ of the order of a few meV (Fig. S7), in agreement with previous findings \textit{(6)}.
Here $k_{\textrm{F}}$ is the Fermi wavevector. We note that $\alpha_{\textrm{R}}$ and $E_{\textrm{SO}}$ are gate-dependent and increase as the conductivity decreases. The variation of spin-orbit coupling with the gate voltage may account for the nonideal linear fits of the conductivities in Fig. 4A,B.

We use these estimates of the Fermi energy and spin-orbit coupling to evaluate the effect of the magnetic field on the Rashba bands, as follows. Consider the Rashba Hamiltonian in the presence of a planar magnetic field

\begin{equation}
    \mathcal{H} = \frac{\hbar^2k^2}{2m} + \alpha_{\textrm{R}}(\mathbf{z}\times\mathbf{k})\cdot\pmb{\sigma} + \beta\mathbf{B}\cdot\pmb{\sigma}.
\end{equation}

\noindent This is the same equation as Eq. \ref{eq:Ham0}, where $\beta$ is the effective magnetic moment and $B$ is in units of Teslas. Its eigenvalues are

\begin{equation}
    E_{\pm} = \frac{\hbar^2k^2}{2m} \pm \sqrt{\alpha_{\textrm{R}}^2k^2 + \beta^2B^2 - 2\alpha_{\textrm{R}}\beta(k_yB_x - k_xB_y)}.
\end{equation}

\noindent The degeneracy point ($E_+ = E_-$) is located at $k^* = \frac{\beta B}{\alpha_{\textrm{R}}}$ and the energy shift of this point caused by the magnetic field is

\begin{equation}
    \Delta E_{\textrm{d}} = E[(k^*(B)] - E[(k^*(0)] = \frac{\hbar^2}{2m}\left(\frac{\beta B}{\alpha_{\textrm{R}}}\right)^2.
    \label{eq:eq_shift}
\end{equation}

\noindent Therefore, the shift of the degeneracy point grows quadratically with the magnetic field. Figure S7C shows this dependence when $\beta$ is taken equal to a Bohr magneton and $\alpha_{\textrm{R}} = 0.3$ meV$\cdot$nm. We find that the energy shift has the same order of magnitude as the Fermi energies of the two Rashba doublets. This calculation confirms, therefore, that magnetic fields of the order of a few T can effectively push the degeneracy point towards and beyond the Fermi level (in agreement with the nonmonotonic dependence of the nonlinear magnetoresistance on the magnetic field). Whether this happens for both Rashba doublets or only for the second band pair depends on the exact values of $E_{\textrm{F}}$ and $\Delta E_{\textrm{d}}$, which are difficult to estimate precisely without going beyond a simple free electron model. In this respect, we also note that $\beta$ is unknown. However, the fact that $E_{\textrm{F}, \, 1} \approx 3E_{\textrm{F}, \, 2}$ and the measured nonlinear response shows only one peak, instead of two, at a given magnetic field (Supplementary Section 6) suggest that the first Rashba doublet contributes little to the nonlinear transport.
\newline

\noindent\underline{6 Magnetic field dependence of the nonlinear longitudinal conductivity}\\

\noindent Figure 3 in the main text shows the second harmonic resistance as a function of the planar magnetic field. However, the quantity that should be compared with the theoretical predictions of the Rashba model (Supplementary Sections 1 and 5) is the nonlinear conductivity, which is shown in Fig. S8. This  conductivity displays a nonmonotonic dependence on the magnetic field. At low field, it grows with odd powers of $B$, namely, $\sim B + B^3 + B^5 +...$. At higher field, instead, it saturates and eventually starts decreasing. This trend is in good agreement with the theoretical prediction shown in Fig. 4 in the main text. A reasonable agreement is also found between the estimated and measured magnetic field $B^*$ at which the nonlinear longitudinal conductivity peaks. The comparison is shown in Fig. S8C. Experimentally, $B^*$ is in the order of 9-12 T. Because the magnetic field at which the nonlinear conductivity is maximum increases beyond the accessible field range as the gate voltage is decreased, $B^*$ can only be evaluated in the high conductivity region. The expected $B^*$ is instead estimated by means of Eq. \ref{eq:eq_shift} for both the first and second Rashba doublets by taking $\Delta E_{\textrm{d}} \approx E_{\textrm{F}}$.

This estimate does not coincide precisely with the experimental $B^*$ because of the uncertainties on $\beta$, which we assume equal to a Bohr magneton, and because of the simplicity of the free electron model, but it is of the correct order of magnitude and reproduces the increase of $B^*$ at lower conductivity. This increase is determined by the rise of $\alpha_{\textrm{R}}$ as the conductivity decreases.
\newline

\noindent\underline{7 Second harmonic longitudinal $B$-symmetric resistance}\\

\noindent In the main text, we discuss the second harmonic longitudinal $B$-antisymmetric resistance because the quantum metric nonlinear conductivity is odd with respect to time reversal and thus vanishes at $B = 0$. The total second harmonic longitudinal resistance includes, though, a $B$-symmetric contribution that is finite at $B = 0$ and originates from side jump and skew scattering mechanisms. As shown in Fig. S9A, the amplitude of $R^{2\omega, \, \textrm{sym}}_{\textrm{yyy}}$ is in general comparable to that of $R^{2\omega, \, \textrm{asym}}_{\textrm{yyy}}$, although with a notable difference. $R^{2\omega, \, \textrm{asym}}_{\textrm{yyy}}$ evolves smoothly with the gate voltage and temperature (see main text). In contrast, $R^{2\omega, \, \textrm{sym}}_{\textrm{yyy}}$ shows a strong enhancement at intermediate voltages and temperatures. This nonmonotonic trend is best seen in the dependence of the nonlinear conductivity $\sigma^{2\omega, \, \textrm{sym}}_{\textrm{yyy}}$ on the zero-field linear conductivity $\sigma_{\textrm{yy}}$, where deviations from the main parabolic-like trend are observed at about $\sigma_{\textrm{yy}} = 5-6$ mS. Although further investigation is required to ascertain the origin of this phenomenon, its occurrence just before and after the second band doublet starts filling suggests a relationship with electronic correlation effects that could possibly activate additional nonlinear scattering mechanisms. Such effects cannot be attributed to the quantum metric, which is odd in the magnetic field. \newline 

\noindent\underline{8 Magnetic field parallel to the $\bar{1}\bar{1}2$ (y) crystallographic direction}\\

\noindent Figure 3 in the main text shows the second harmonic longitudinal $B$-antysimmetric resistance when the magnetic field is oriented along the $\bar{1}10$ $(x)$ crystallographic direction. In Fig. S10, we show that a qualitatively similar response is obtained upon reorienting the magnetic field along the $\bar{1}\bar{1}2$ $(y)$ direction. Also in this case the nonlinear response is enhanced (suppressed) when $\mathbf{B}\perp\mathbf{I}$ ($\mathbf{B}\parallel\mathbf{I}$). The nonzero resistance measured when $\mathbf{B}\parallel\mathbf{I}$ is attributed to the combination of the trigonal warping with a possible misalignment of the magnetic field. This phenomenology confirms the model in Supplementary Section 1. Note that the measurements with the current flowing along the $x$ and $y$ directions are performed in two physically distinct Hall bar devices. Therefore, the presence of the quantum metric and bilinear magnetoresistance in different devices confirms the reproducibility of our findings.

The nonlinear conductivity $\sigma^{2\omega}_{\textrm{xxx}}$ calculated from $R^{2\omega, \, \textrm{asym}}_{\textrm{xxx}}$ is shown in Fig. S10C. Comparing this figure with Fig. 3 in the main text demonstrates that $R^{2\omega, \, \textrm{asym}}_{\textrm{xxx}}$ appears simultaneously to the filling of the second Rashba doublet. Moreover, the linear dependence of $\sigma^{2\omega}_{\textrm{xxx}}$ on the square mobility $\mu_2^2$ confirms the coexistence of the quantum metric transport and bilinear magnetoresistance, as already deduced for $\sigma^{2\omega}_{\textrm{yyy}}$. These two contributions increase initially with the magnetic field and saturate thereafter.
\newline

\noindent\underline{9 Temperature dependence of the linear and nonlinear transport}\\

\noindent In the main text, we focus on the linear and nonlinear transport at fixed temperature and variable gate voltage. Here, we discuss the transport measurements at fixed gate voltage $V_{\textrm{g}} = 20 $ V and variable temperature (Fig. S11). In general, temperature has an effect qualitatively similar to that of the gate voltage. The first harmonic transverse and second harmonic longitudinal resistances share the same phenomenology as the corresponding signals measured at different gate voltages. In particular, both the anomalous planar Hall effect and the second harmonic longitudinal resistance increase as temperature is decreased. Moreover, the second harmonic longitudinal response is enhanced (suppressed) when $\mathbf{B}\perp\mathbf{I}$ ($\mathbf{B}\parallel\mathbf{I}$). The comparison between the nonlinear conductivity and the carrier density suggests also in this case the dominant role of the second Rashba doublet. The dependence of the nonlinear conductivity on the square of the electronic mobility, however, is not linear as expected from the scaling relation $\sigma^{2\omega} = \sigma_{\textrm{qm}} + \sigma_{\textrm{bm}}\mu_2^2$. This deviation from the predicted linear trend likely has the same origin as the milder nonlinearity observed in Figs. 4A and 4B in the main text, namely, the change of the electronic effective mass and spin-orbit coupling with temperature. The temperature dependence of these two parameters is, in general, an open point of the physics of the 2D electron gas at LaAlO$_3$/SrTiO$_3$ interfaces. Previous work has shown that varying the temperature leads to a renormalization of the chemical potential that is qualitatively similar to the chemical potential tuning caused by the back gate voltage (\textit{39}). This is confirmed by the disappearance of the second electronic population at temperatures above 12-14 K (Fig. S11E). However, how temperature influences quantitatively the effective electronic mass and the spin-orbit interaction remains unclear. Given this uncertainties, we show in Fig. S11F a tentative linear fit of the nonlinear conductivity to the nominal trend $\sigma^{2\omega} = \sigma_{\textrm{qm}} + \sigma_{\textrm{bm}}\mu_2^2$ at temperatures below 7 K. These approximated fits extrapolate to a nonzero intercept with the $y$ axis, which suggests the presence of an intrinsic contribution, i.e., the quantum metric (Fig. S11G), in agreement with the dataset and analysis reported in the main text as a function of the gate voltage. A poorer but more inclusive fit to the entire dataset, i.e., all temperatures, leads to the same magnetic field dependence of the quantum metric as in Fig. S11G but with about halved magnitude.
\newline

\noindent\underline{10 Exclusion of spurious effects}\\

\noindent Here we exclude that Joule heating or other spurious effects can be mistaken for the nonlinear transport induced by the quantum metric.

\begin{itemize}
    \item In principle, since Joule heating scales with the square of the current, thermal gradients can induce a nonlinear response. Of these thermal gradients, only that perpendicular to the sample surface could generate a nonlinear response because in-plane gradients, if any, are spatially symmetric and cannot generate finite transverse or longitudinal signals. Now, a perpendicular thermal gradient can cause the anomalous Nernst effect and the Seebeck effect. The latter, however, is even with respect to time reversal and cannot therefore contribute to the $B$-antisymmetric signals that we analyze to extract the quantum metric. Moreover, the Seebeck effect generates a signal parallel to the thermal gradient and cannot contribute to the transverse and longitudinal voltages, which are planar signals. 
    The anomalous Nernst effect cannot either explain our experimental observations for four reasons. First, the second harmonic voltages that we measure, either $B$-symmetric or $B$-antisymmetric, are strongly nonmonotonic with respect to the magnetic field, which is not compatible with a possible magnetization being saturated by the field. Second, the anomalous Nernst effect is expected to give rise to identical second harmonic transverse and longitudinal voltages (up to an in-plane rotation of the magnetic field by 90°). In such a case, the ratio between the transverse and longitudinal resistances should equal the ratio $W/L = 1/3$ between the device width and length, which is not our case. 
    Third, the anomalous Nernst effect is not expected to show such a strong dependence on temperature and gate voltage as we observe experimentally. Fourth, the anomalous Nernst current scales as $J_{\textrm{N}} = \sigma_{\textrm{N}}\Delta T \sim \sigma_{\textrm{N}}I_{\omega}^2R_{\textrm{ii}}^{\omega} \sim \sigma_{\textrm{N}}E_{\omega}^2R_{\textrm{ii}}^{\omega}$. However, since the Nernst conductivity $\sigma_{\textrm{N}}$ is independent of the scattering time $\tau$ while $I_{\omega}^2R_{\textrm{ii}}^{\omega}\sim \tau$ \textit{(26)},
    the anomalous Nernst conductivity scales linearly with $\tau$.
    
    \item The nonlinear response that we measure cannot originate from nonohmic contacts or Schottky barriers. First, as shown in Fig. S12A, the two-probe current-voltage characteristic is linear in the entire current range, which demonstrates the ohmic nature of the electric contacts. Note that all the data presented in the main text and Supplementary Sections were measured with a current of 50 \textmu A. Second, 
    such a spurious effect should be independent of the gate voltage and temperature, in contrast to the phenomenology of our measurements. Third, nonohmic contacts cannot explain the nonmonotonic dependence of the nonlinear signals on the magnetic field.

    \item The nonlinear response cannot originate from capacitive or inductive effects either because: 1) the second harmonic response is independent of the selected frequency over more than a decade (Fig. S12C); 2) capacitive and inductive effects can hardly explain the temperature, gate voltage, field amplitude, and field direction dependence of the measured signals.

    \item As pointed out in Ref. \textit{(59)}, spurious nonlinear transverse and longitudinal signals can be caused by the modulation of the (fixed) DC back gate voltage by the AC current used to measure the second harmonic response. Because the sample properties (carrier density and electronic mobility, hence magnetoresistance) depend strongly on the gate voltage, the modulation of the latter results in nonlinear transport effects that can mimic the nonlinear signal driven by the quantum geometry. This artefact can be identified by comparing signals measured in different grounding conditions because changing the position of the grounded terminal(s) leads to a redistribution of the electric potential across the device and, consequently, of the effective gate voltage \textit{(59)}. We exploit this approach to exclude the possibility of artefacts. As shown in Fig. S13, we consider two contact configurations that differ in the position of the positive and grounded electrodes used to inject the AC current. The sensing contacts used to probe linear and nonlinear voltages are instead unchanged. Note that we take the absolute value of the linear longitudinal and Hall resistances because inverting the current electrodes while keeping fixed the sensing leads results in the sign inversion of the measured linear voltages. The sign of the nonlinear signal remains instead the same, which is consistent with its second harmonic origin.
    
    Upon exchanging the current electrodes, we find that the sample resistance shows a small but finite change that indicates a variation of the sample properties caused by the modification of the potential landscape (Fig. S13B). This change is indeed accompanied by a variation of both the linear Hall resistance and nonlinear longitudinal resistance (Fig. S13C-D). However, symmetrizing and antisymmetrizing the data with respect to the magnetic field shows that only the B-symmetric component is affected by the redistribution of the electric potential while the B-antisymmetric component is left unchanged. This evidence proves that both the anomalous planar Hall effect and the B-antisymmetric nonlinear longitudinal resistance (associated with the Berry curvature and quantum metric, respectively) are not artefacts but reflect the quantum geometrical properties of the bandstructure.

    Additional considerations support this conclusion. First, we note that the nonlinear longitudinal resistance is the largest at high gate voltage and low temperature, i.e., when the linear resistance is minimum. Upon decreasing the gate voltage or increasing the temperature, the nonlinear response disappears while the sample resistance increases. These opposite trends rule out the scenario considered in Ref. \textit{(59)} according to which we should expect that the nonlinear spurious resistance induced by the gate voltage modulation increases with the linear resistance. Second, the nonlinear resistance that we measure is maximum when the electric current and planar magnetic field are collinear and is strongly suppressed when they are orthogonal. This dependence cannot be explained in the scenario of Ref. \textit{(59)}.
    
\end{itemize}

\noindent We further note that

\begin{itemize}
    \item The second harmonic voltage scales quadratically with the applied current up to the maximum value (100 \textmu A), as shown in Fig. S12B and as expected for a second order effect.
    \item The second harmonic response is entirely out of phase with respect to the injected current (Fig. S12D-E), consistent with a second-order effect. The first harmonic resistance is instead in phase.
\end{itemize}

\clearpage

\begin{figure*}[t]
\centering
\includegraphics{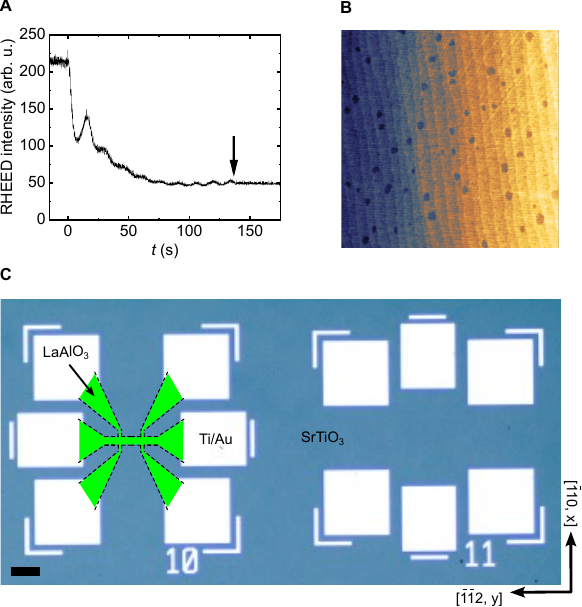}
\caption{\textbf{Fig. S1. Sample growth.} (\textbf{A}) Oscillations of the RHEED intensity during the pulsed laser deposition of 9 unit cells of LaAlO$_3$ on the 111-oriented SrTiO$_3$ substrate. The arrow indicates the end of the growth. (\textbf{B}) Atomic force microscopy of the terraced surface of SrTiO$_3$/LaAlO$_3$. Note that the holes in the terraces were already present in the purchased SrTiO$_3$ substrates. The image size is 3x3 \textmu m$^2$. (\textbf{C}) Optical image of two patterned Hall bar devices oriented along orthogonal directions on the SrTiO$_3$ surface. The green area indicates the position and shape of the devices. Note that the LaAlO$_3$ layer cannot be distinguished by eye from the substrate. The black scale bar corresponds to 60 \textmu m.} 
\label{figS1}
\end{figure*}

\clearpage

\begin{figure*}[t]
    \begin{center}
    \includegraphics[width=0.99\textwidth]{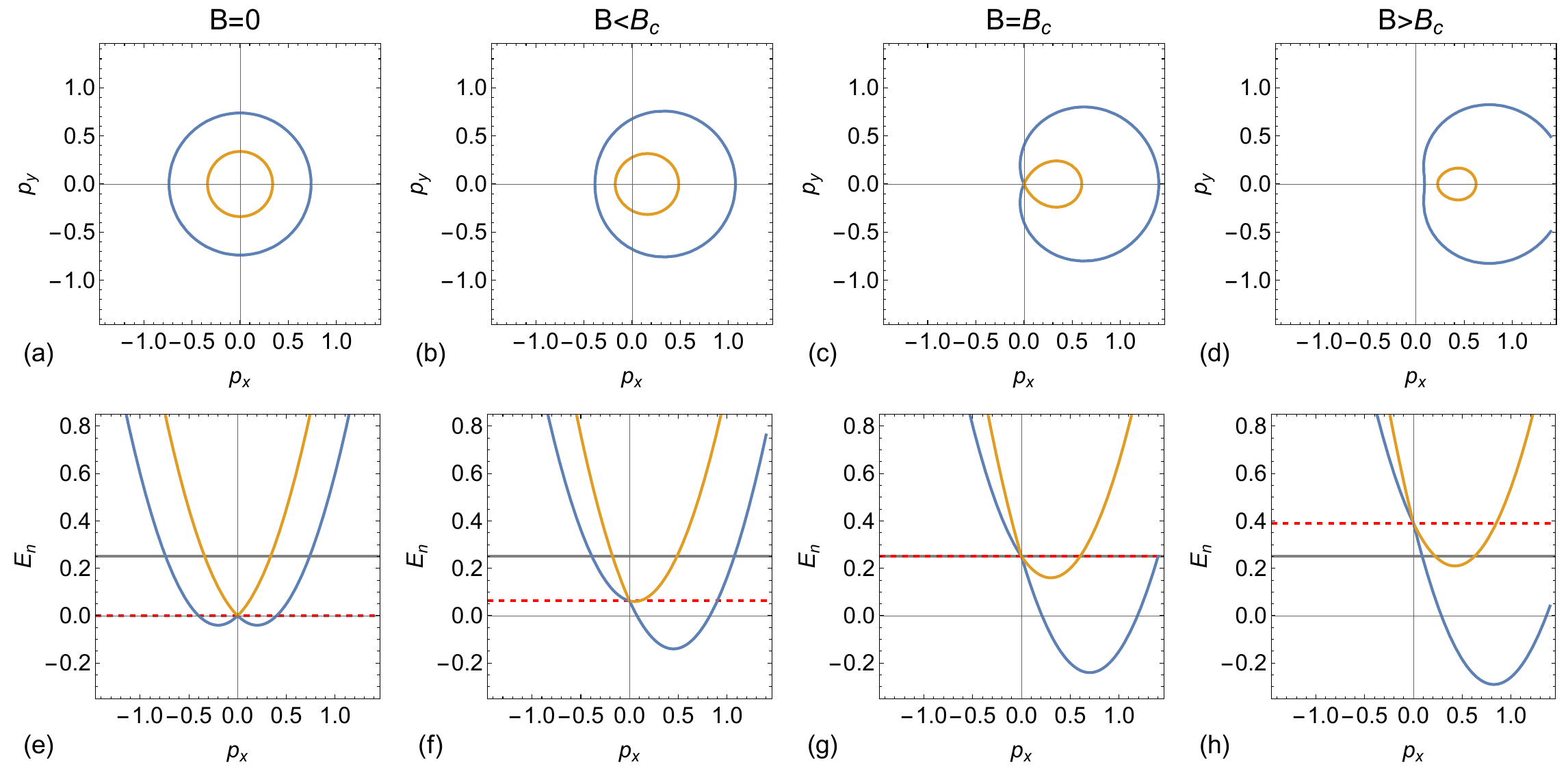}
    \caption{\textbf{Fig. S2. Rashba bands in the presence of a planar magnetic field.} (\textbf{A-D}) Fermi lines for $\alpha_R=0.4 E_0/k_F^0$, $\mu=0.25E_0$ and for 4 different values of magnetic field $B$ (namely $B=0$; $B=0.1 E_0$, $B=0.2 E_0$ and  $B=0.25 E_0$) placed along the $y$ direction ($\theta=\pi/2$) in the shifted-wave-vector space $(p_x,p_y)$. (\textbf{E-H}) Energy bands $E_\pm$ for the same set of parameters as a function of $p_x$ with $p_y=0$. The red dashed line marks the energy of the crossing poing ($E_c=B^2/\alpha_R^2$) while the gray line indicates the level of the chemical potential $\mu$ used for the Fermi lines in plots \textbf{A-D}.}
    \label{fig:EnergyFL}
    \end{center}
\end{figure*}

\clearpage

\begin{figure*}[t]
    \begin{center}
    \includegraphics[width=0.98\textwidth]{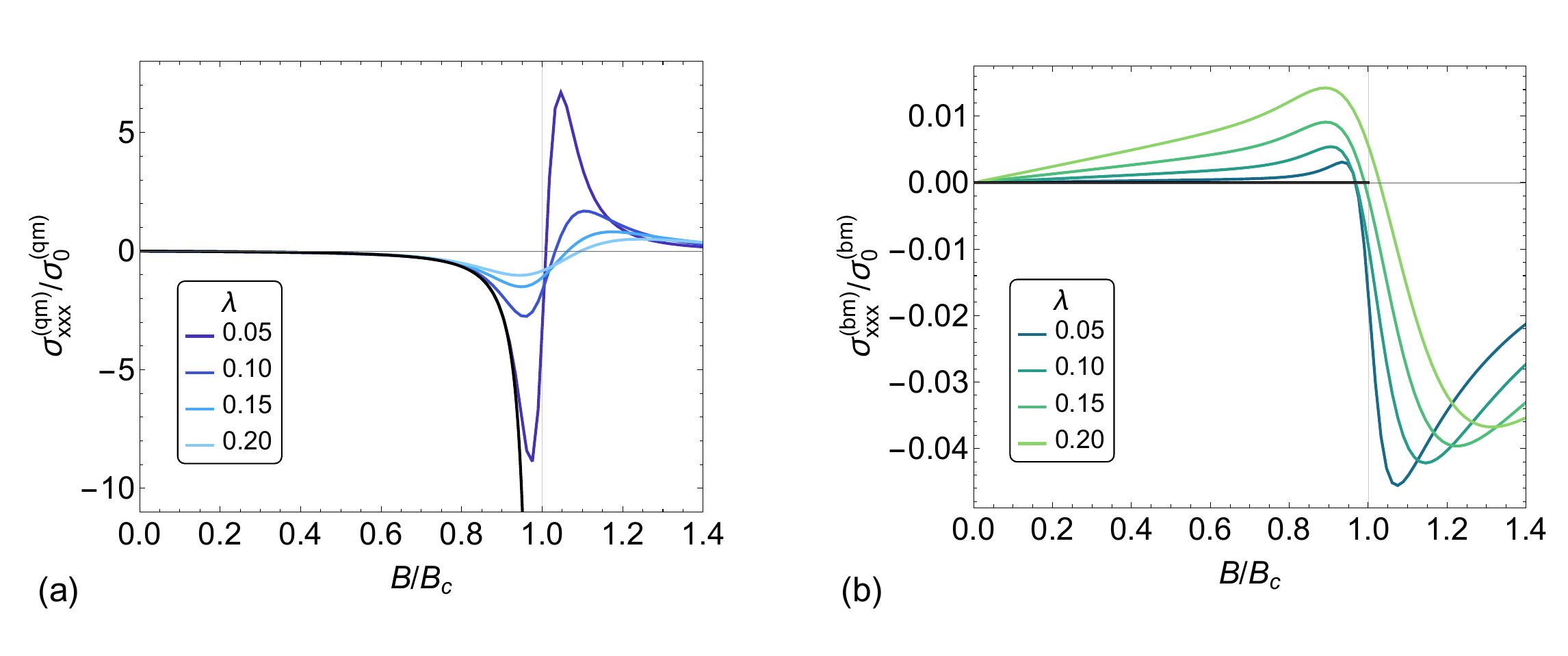}
    \caption{\textbf{Fig. S3. Quantum metric and bilinear magnetoelectric conductivities.} Plot of the quantum metric contribution to the nonlinear conductivity $\sigma_{xxx}^{(\rm{qm})}$ (\textbf{A}) and the bilinear magnetoelectric resistance contribution $\sigma_{xxx}^{(\rm{bm})}$ (\textbf{B}) for $\alpha_R=0.4 E_0/k_F^0$, $\mu=0.5E_0$ and four different values of warping parameter $\lambda$ as a function of the magnetic field $B$ oriented along the $y$ direction. The black line shows the analytical results for $\lambda=0$ up to the critical field $B=B_{\textrm{c}}$.}
    \label{fig:theory}
    \end{center}
\end{figure*}

\clearpage

\begin{figure*}[t]
\centering
\includegraphics{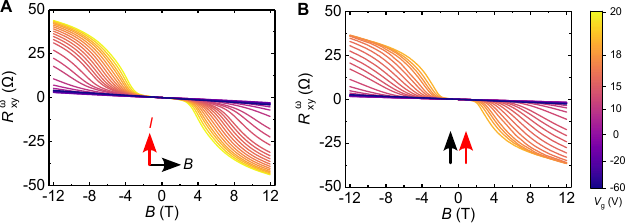}
\caption{\textbf{Fig. S4. Planar Hall effect.} (\textbf{A-B}) First harmonic transverse resistance as a function of the magnetic field $B$ and gate voltage $V_{\textrm{g}}$ for two relative orientations of the magnetic field (black arrow) and electric current (red arrow) at a temperature of 3 K. The electric current is sourced along the $\bar{1}\bar{1}2$ $(y)$ crystallographic direction.} 
\label{figS7}
\end{figure*}

\clearpage

\begin{figure*}[t]
\centering
\includegraphics{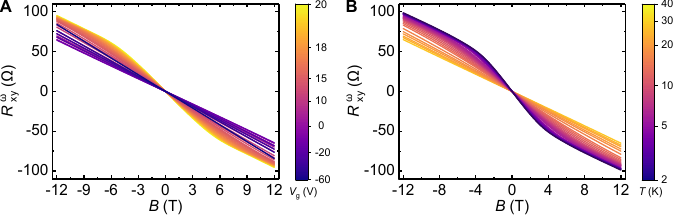}
\caption{\textbf{Fig. S5. Ordinary Hall effect.} (\textbf{A}) First harmonic transverse resistance measured while sweeping the magnetic field along the direction normal to the sample surface at variable gate voltage and temperature $T = 3$ K. (\textbf{B}) The same as \textbf{A} at different temperatures and gate voltage $V_{\textrm{g}} = 20$ V.} 
\label{figS14}
\end{figure*}

\clearpage

\begin{figure*}[t]
\centering
\includegraphics{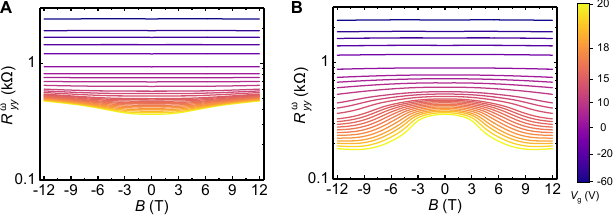}
\caption{\textbf{Fig. S6. Linear magnetoresistance.} (\textbf{A}) Log-scale first harmonic longitudinal resistance as a function of the magnetic field oriented along the surface normal at variable gate voltage and temperature $T = 3$ K. (\textbf{B}) The same as \textbf{A} as a function of the magnetic field oriented in the sample plane.} 
\label{figS8}
\end{figure*}

\clearpage

\begin{figure*}[t]
\centering
\includegraphics{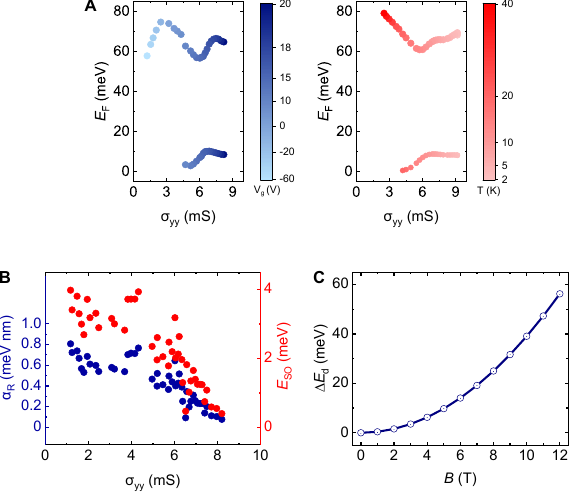}
\caption{\textbf{Fig. S7. Energy scales of 2D electron gas at the 111-oriented LaAlO$_3$/SrTiO$_3$ interface.} (\textbf{A}) Fermi energy of the two Rashba doublets as a function of the gate voltage at a temperature $T = 3$ K (left) and as a function of temperature at a gate voltage $V_{\textrm{g}} = 20$ V (right). (\textbf{B}) Rashba parameter (left axis) and Rashba splitting (right axis) as a function of the gate-modulated longitudinal conductivity. (\textbf{C}) Energy shift of the degeneracy point of the Rasha bands as a function of the planar magnetic field.} 
\label{figS13}
\end{figure*}

\clearpage

\begin{figure*}[t]
\centering
\includegraphics{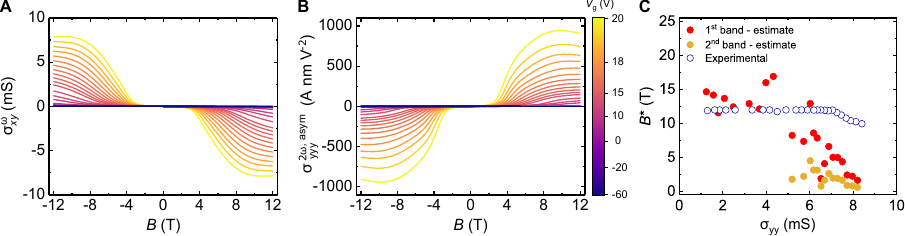}
\caption{\textbf{Fig. S8. Magnetic field dependence of the nonlinear longitudinal conductivity.} (\textbf{A}) Anomalous planar Hall conductivity as a function of the magnetic field $B$ and gate voltage $V_{\textrm{g}}$ at a temperature $T = 3$ K. (\textbf{B}) The same as \textbf{A} for the nonlinear longitudinal conductivity. (\textbf{C}) Comparison between the observed and estimated magnetic field $B^*$ at which the nonlinear longitudinal conductivity reaches its maximum. Experimentally, $B^*$ is obtained from the data in \textbf{B}.} 
\label{figS4}
\end{figure*}

\clearpage

\begin{figure*}[t]
\centering
\includegraphics{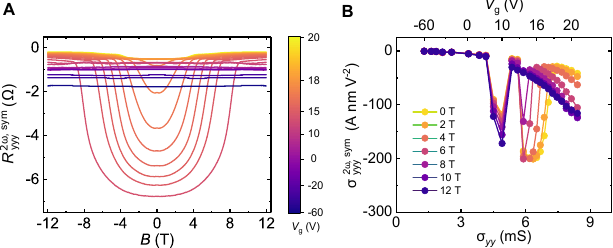}
\caption{\textbf{Fig. S9. Second harmonic longitudinal $B$-symmetric resistance.} (\textbf{A}) Second harmonic longitudinal $B$-symmetric resistance measured as a function of the magnetic field $B$ and gate voltage $V_{\textrm{g}}$ when $\mathbf{B}\perp\mathbf{I}(\parallel\mathbf{x})$ at a temperature $T = 3$ K. (\textbf{B}) Nonlinear conductivity calculated from \textbf{A} at different magnetic fields as a function of the zero-field linear conductivity.} 
\label{figS2}
\end{figure*}

\clearpage

\begin{figure*}[t]
\centering
\includegraphics{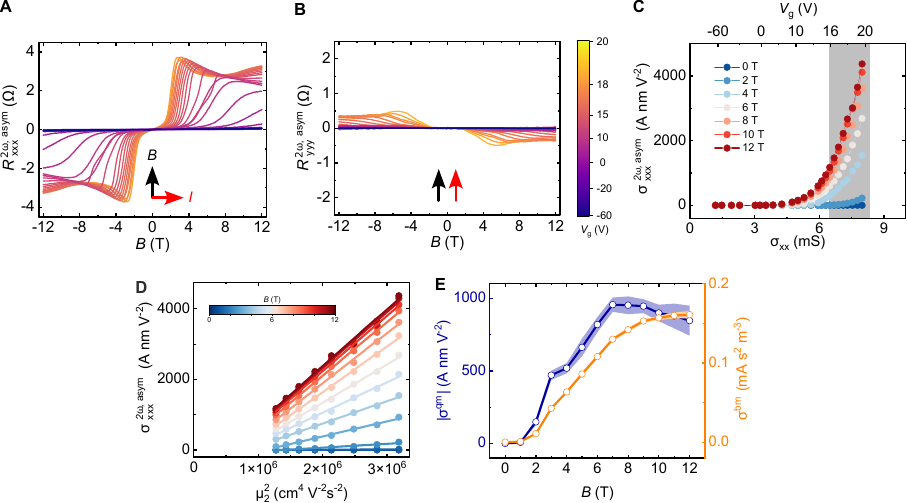}
\caption{\textbf{Fig. S10. Nonlinear electronic transport with $\mathbf{B}\parallel\mathbf{y}$.} (\textbf{A-B}) Second harmonic longitudinal $B$-antisymmetric resistance measured as a function of the magnetic field $B$ and gate voltage $V_{\textrm{g}}$ for two relative orientations of $\mathbf{B}\parallel\mathbf{y}$ (black arrow) and the electric current $I$ (red arrow) at a temperature $T = 3$ K. The electric current is applied along the $\bar{1}10$ (x) and $\bar{1}\bar{1}2$ (y) crystallographic directions in \textbf{A} and \textbf{B}, respectively. (\textbf{C}) Nonlinear longitudinal conductivity calculated from  \textbf{A} as a function of the linear conductivity. (\textbf{D}) Dependence of the nonlinear conductivity on the square of the mobility of the second Rashba pair. The lines are linear fits to the data. (\textbf{E}) Magnetic field dependence of the quantum metric and bilinear magnetoresistance contributions to the total nonlinear conductivity obtained from the fits in \textbf{D}. The shaded areas correspond to the error of the fits.} 
\label{figS3}
\end{figure*}

\clearpage

\begin{figure*}[t]
\centering
\includegraphics{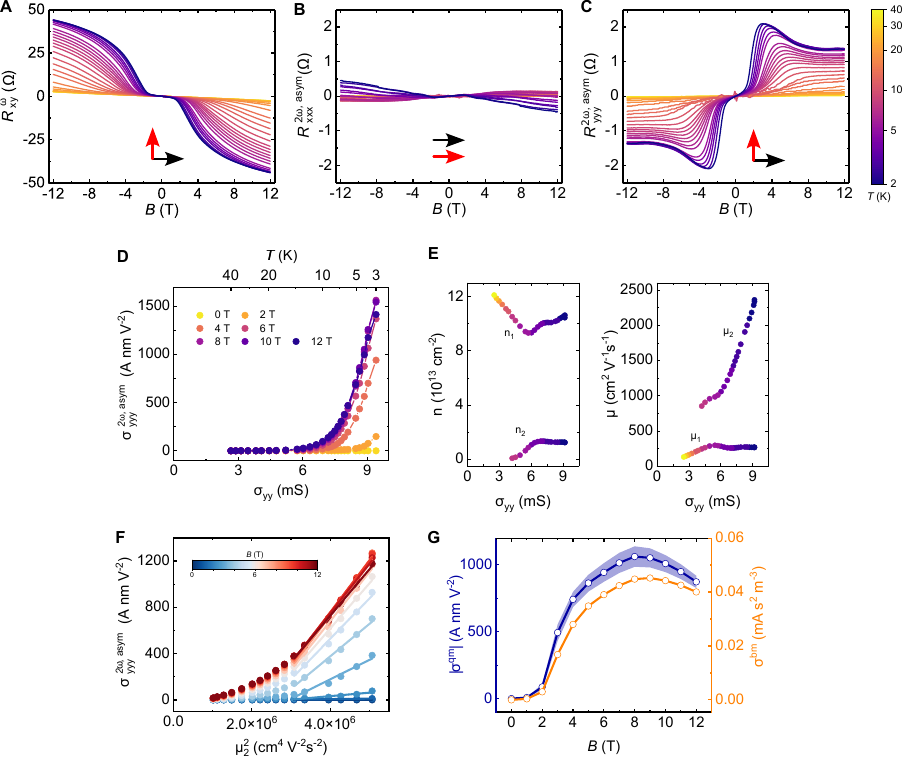}
\caption{\textbf{Fig. S11. Nonlinear transport at variable temperature.} (\textbf{A-C}) First harmonic transverse and second harmonic longitudinal resistances as a function of the in-plane magnetic field $B$ and temperature at a gate voltage $V_{\textrm{g}} = 20 $ V.  The black and red arrows indicate the direction of $B$ and the electric current $I$, respectively. The current is applied along the $\bar{1}10$ $(x)$ and $\bar{1}\bar{1}2$ $(y)$ crystallographic directions in \textbf{B} and \textbf{A, C}, respectively. (\textbf{D}) Nonlinear longitudinal conductivity calculated from \textbf{C} at different magnetic fields as a function of the zero-field longitudinal linear conductivity. (\textbf{E}) Two-band electron densities $n$ and mobilities $\mu$ at variable temperature and a gate voltage $V_{\textrm{g}} = 20 $ V. (\textbf{F}) Dependence of the nonlinear conductivity on the square of the electron mobility of the second Rashba pair.}  
\label{figS10}
\end{figure*}

\clearpage

\begin{figure*}[t]
\centering
\includegraphics{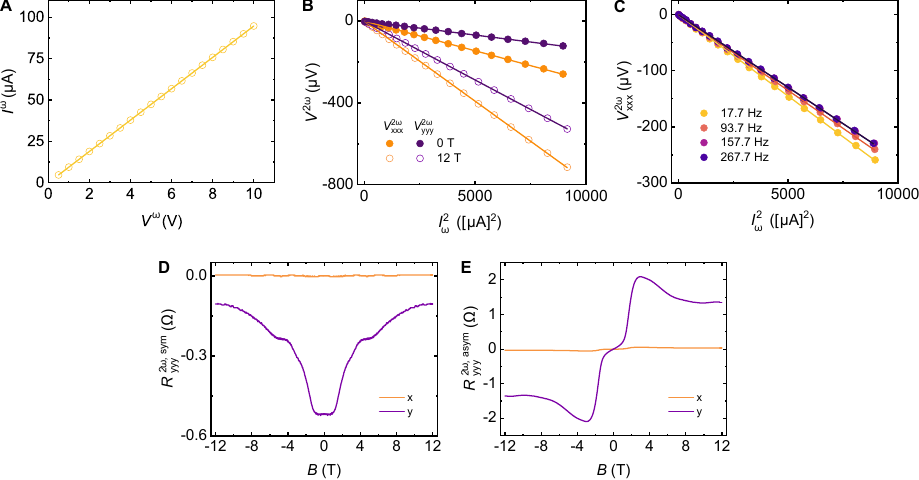}
\caption{\textbf{Fig. S12. Current, frequency, and phase depedence of the nonlinear transport.} (\textbf{A}) Dependence of the injected current on the two-probe applied voltage. (\textbf{B}) Dependence of the measured second harmonic longitudinal voltage on the square of the injected current at a magnetic field of 0 T and 12 T, at a frequency of $\frac{\omega}{2\pi} = 17.7$ Hz, and for two directions of the current. (\textbf{C}) Dependence of the measured second harmonic longitudinal voltage on the square of the injected current at different frequencies and at a magnetic field of 0 T. The lines in \textbf{A-C} are linear fits to the data with zero intercept. (\textbf{D-E}) Comparison of the in-phase $(x)$ and out-of-phase $(y)$ components of the second harmonic longitudinal $B$-symmetric and $B$-antisymmetric responses, respectively.} 
\label{figS6}
\end{figure*}

\begin{figure*}[t]
\centering
\includegraphics{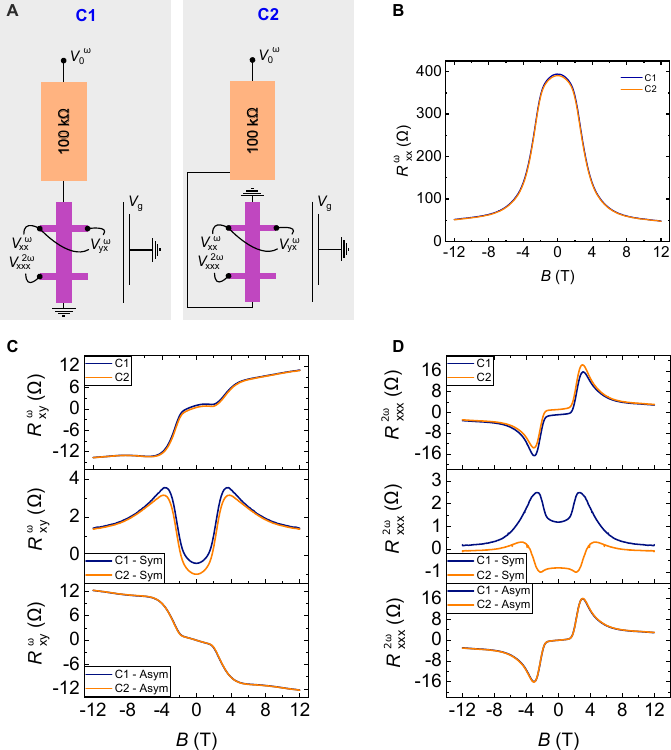}
\caption{\textbf{Fig. S13. Grounding effects.} (\textbf{A}) Contact configurations used to inject an AC current and measure linear and nonlinear voltages. In configuration 2 (C2) the positive and grounded electrodes used to inject the current are inverted as compared to configuration 1 (C1). (\textbf{B}) Linear magnetoresistance measured in the two configurations. (\textbf{C}) Linear Hall resistance measured in the two configurations. The upper panel shows the raw data while the middle (bottom) panel shows the B-symmetric (B-antisymmetric) component obtained by symmetrizing and antisymmetrizing the data with respect to the magnetic field. (\textbf{D}) The same as \textbf{C} for the nonlinear longitudinal resistance. All data were acquired while sweeping the magnetic field in the sample plane along the $y$ direction at a temperature of 3 K, gate voltage of 19 V, and injected current of 80 \textmu A.} 
\label{figS6}
\end{figure*}


\begin{thebibliography}{10}

\bibitem{tokura2018}
Y.~Tokura, N.~Nagaosa, {\it Nat. Commun.\/} {\bf 9}, 3740 (2018).

\bibitem{Xiao2010}
D.~Xiao, M.-C. Chang, Q.~Niu, {\it Rev. Mod. Phys.\/} {\bf 82}, 1959 (2010).

\bibitem{Nagaosa2010}
N.~Nagaosa, J.~Sinova, S.~Onoda, A.~H. MacDonald, N.~P. Ong, {\it Rev. Mod.
  Phys.\/} {\bf 82}, 1539 (2010).

\bibitem{Sodemann2015}
I.~Sodemann, L.~Fu, {\it Phys. Rev. Lett.\/} {\bf 115}, 216806 (2015).

\bibitem{Ma2019a}
Q.~Ma, {\it et~al.\/}, {\it Nature\/} {\bf 565}, 337 (2019).

\bibitem{Lesne2022a}
E.~Lesne, {\it et~al.\/}, {\it Nat. Mater.\/} {\bf 22}, 576 (2023).

\bibitem{Du2021}
Z.~Z. Du, H.-Z. Lu, X.~C. Xie, {\it Nat. Rev. Phys.\/} {\bf 3}, 744 (2021).

\bibitem{Ortix2021}
C.~Ortix, {\it Adv. Quantum Technol.\/} {\bf 4}, 1 (2021).

\bibitem{Makushko2023}
P.~Makushko, {\it et~al.\/}, {\it Nat. Electron.\/} {\bf 7}, 207 (2024).

\bibitem{Kaplan2022}
D.~Kaplan, T.~Holder, B.~Yan, {\it Phys. Rev. Lett.\/} {\bf 132}, 026301
  (2024).

\bibitem{Lahiri2023}
K.~Das, S.~Lahiri, R.~B. Atencia, D.~Culcer, A.~Agarwal, {\it Phys. Rev. B\/}
  {\bf 108}, L201405 (2023).

\bibitem{Provost1980}
J.~P. Provost, G.~Vallee, {\it Commun. Math. Phys.\/} {\bf 76}, 289 (1980).

\bibitem{Cheng2010}
R.~Cheng, {\it arXiv 1012.1337\/} pp. 1--9 (2010).

\bibitem{Torma2023}
P.~Torma, {\it Phys. Rev. Lett.\/} {\bf 131}, 240001 (2023).

\bibitem{Resta2011}
R.~Resta, {\it Eur. Phys. J. B\/} {\bf 79}, 121 (2011).

\bibitem{Peotta2015}
S.~Peotta, P.~T{\"{o}}rm{\"{a}}, {\it Nat. Commun.\/} {\bf 6}, 8944 (2015).

\bibitem{Rossi2021}
E.~Rossi, {\it Curr. Opin. Solid State Mater. Sci.\/} {\bf 25}, 100952 (2021).

\bibitem{Huhtinen2022}
K.-E. Huhtinen, J.~Herzog-Arbeitman, A.~Chew, B.~A. Bernevig,
  P.~T{\"{o}}rm{\"{a}}, {\it Phys. Rev. B\/} {\bf 106}, 014518 (2022).

\bibitem{Tan2019a}
X.~Tan, {\it et~al.\/}, {\it Phys. Rev. Lett.\/} {\bf 122}, 210401 (2019).

\bibitem{Gianfrate2020}
A.~Gianfrate, {\it et~al.\/}, {\it Nature\/} {\bf 578}, 381 (2020).

\bibitem{Yu2020}
M.~Yu, {\it et~al.\/}, {\it Natl. Sci. Rev.\/} {\bf 7}, 254 (2020).

\bibitem{Ren2021}
J.~Ren, {\it et~al.\/}, {\it Nat. Commun.\/} {\bf 12}, 689 (2021).

\bibitem{Liao2021}
Q.~Liao, {\it et~al.\/}, {\it Phys. Rev. Lett.\/} {\bf 127}, 107402 (2021).

\bibitem{Tian2023}
H.~Tian, {\it et~al.\/}, {\it Nature\/} {\bf 614}, 440 (2023).

\bibitem{Yi2023}
C.-R. Yi, {\it et~al.\/}, {\it Phys. Rev. Res.\/} {\bf 5}, L032016 (2023).

\bibitem{Gao2023}
A.~Gao, {\it et~al.\/}, {\it Science (80-. ).\/} {\bf 381}, 181 (2023).

\bibitem{Wang2023c}
N.~Wang, {\it et~al.\/}, {\it Nature\/} {\bf 621}, 487 (2023).

\bibitem{Han2024a}
J.~Han, {\it et~al.\/}, {\it Nat. Phys.\/} {\bf 20}, 1110 (2024).

\bibitem{Li2014}
C.~H. Li, {\it et~al.\/}, {\it Nat. Nanotechnol.\/} {\bf 9}, 218 (2014).

\bibitem{Bihlmayer2022}
G.~Bihlmayer, P.~No{\"{e}}l, D.~V. Vyalikh, E.~V. Chulkov, A.~Manchon, {\it
  Nat. Rev. Phys.\/} {\bf 4}, 642 (2022).

\bibitem{Manipatruni2019}
S.~Manipatruni, {\it et~al.\/}, {\it Nature\/} {\bf 565}, 35 (2019).

\bibitem{Noel2020}
P.~No{\"{e}}l, {\it et~al.\/}, {\it Nature\/} {\bf 580}, 483 (2020).

\bibitem{suppScience}
{Additional data and analyses are available as supplementary materials}.

\bibitem{He2018a}
P.~He, {\it et~al.\/}, {\it Nat. Phys.\/} {\bf 14}, 495 (2018).

\bibitem{Tuvia2023}
G.~Tuvia, {\it et~al.\/}, {\it Phys. Rev. Lett.\/} {\bf 132}, 146301 (2024).

\bibitem{Caviglia2008}
A.~D. Caviglia, {\it et~al.\/}, {\it Nature\/} {\bf 456}, 624 (2008).

\bibitem{Monteiro2019}
A.~M. Monteiro, {\it et~al.\/}, {\it Phys. Rev. B\/} {\bf 99}, 201102 (2019).

\bibitem{Rodel2014}
T.~C. R{\"{o}}del, {\it et~al.\/}, {\it Phys. Rev. Appl.\/} {\bf 1}, 1 (2014).

\bibitem{Khanna2019}
U.~Khanna, {\it et~al.\/}, {\it Phys. Rev. Lett.\/} {\bf 123}, 036805 (2019).

\bibitem{Diez2015}
M.~Diez, {\it et~al.\/}, {\it Phys. Rev. Lett.\/} {\bf 115}, 016803 (2015).

\bibitem{Battilomo2021}
R.~Battilomo, N.~Scopigno, C.~Ortix, {\it Phys. Rev. Res.\/} {\bf 3}, L012006
  (2021).

\bibitem{Trama2022b}
M.~Trama, V.~Cataudella, C.~A. Perroni, F.~Romeo, R.~Citro, {\it Phys. Rev.
  B\/} {\bf 106}, 075430 (2022).

\bibitem{Du2019}
Z.~Z. Du, C.~M. Wang, S.~Li, H.-Z. Lu, X.~C. Xie, {\it Nat. Commun.\/} {\bf
  10}, 3047 (2019).

\bibitem{Borunda2007}
M.~Borunda, {\it et~al.\/}, {\it Phys. Rev. Lett.\/} {\bf 99}, 066604 (2007).

\bibitem{Ohtomo2004a}
A.~Ohtomo, H.~Y. Hwang, {\it Nature\/} {\bf 427}, 423 (2004).

\bibitem{Vicente-Arche2021}
L.~M. Vicente‐Arche, {\it et~al.\/}, {\it Adv. Mater.\/} {\bf 33}, 1 (2021).

\bibitem{Nitta1997}
J.~Nitta, T.~Akazaki, H.~Takayanagi, T.~Enoki, {\it Phys. Rev. Lett.\/} {\bf
  78}, 1335 (1997).

\bibitem{Ideue2017}
T.~Ideue, {\it et~al.\/}, {\it Nat. Phys.\/} {\bf 13}, 578 (2017).

\bibitem{Li2021e}
Y.~Li, {\it et~al.\/}, {\it Nat. Commun.\/} {\bf 12}, 540 (2021).

\bibitem{Ast2007}
C.~R. Ast, {\it et~al.\/}, {\it Phys. Rev. Lett.\/} {\bf 98}, 186807 (2007).

\bibitem{Sanchez2013}
J.~C.~R. S{\'{a}}nchez, {\it et~al.\/}, {\it Nat. Commun.\/} {\bf 4}, 2944
  (2013).

\bibitem{Ahn2020}
J.~Ahn, G.-Y. Guo, N.~Nagaosa, {\it Phys. Rev. X\/} {\bf 10}, 041041 (2020).

\bibitem{Ma2023a}
Q.~Ma, R.~{Krishna Kumar}, S.-Y. Xu, F.~H.~L. Koppens, J.~C.~W. Song, {\it Nat.
  Rev. Phys.\/} {\bf 5}, 170 (2023).

\bibitem{Feng2024}
X.~Feng, {\it et~al.\/}, {\it Mater. Today Quantum\/} {\bf 6}, 100040 (2025).

\bibitem{zenodo_Science2024}
G.~Sala, {The quantum metric of electrons with spin-momentum locking,
  https://doi.org/10.5281/zenodo.10692339} (2024).

\bibitem{Bures1969}
D.~Bures, {\it Trans. Am. Math. Soc.\/} {\bf 135}, 199 (1969).

\bibitem{Hikami1980}
S.~Hikami, A.~I. Larkin, Y.~Nagaoka, {\it Prog. Theor. Phys.\/} {\bf 63}, 707
  (1980).

\bibitem{Maekawa1981}
S.~Maekawa, H.~Fukuyama, {\it J. Phys. Soc. Japan\/} {\bf 50}, 2516 (1981).

\bibitem{Xu2024z}
L.~Xu, {\it et~al.\/}, {\it npj Quantum Mater.\/} {\bf 9}, 79 (2024).

\end{thebibliography}
\end{document}